\documentclass[acmsmall]{acmart}
\AtBeginDocument{%
  }

\usepackage{graphicx}
\usepackage{caption}
\usepackage{subcaption}
\usepackage{bbold}
\usepackage{mathtools}
\usepackage{todonotes}
\presetkeys%
    {todonotes}%
    {inline,backgroundcolor=yellow}{}

\setcopyright{acmcopyright}
\copyrightyear{2025}
\acmYear{2025}
\acmJournal{TORS}
\acmVolume{1}
\acmNumber{1}
\acmArticle{1}
\acmMonth{1}

\usepackage{xspace}
\newcommand{\initialNbOfPapers}[0]{51 }
\newcommand{\finalNbOfPapers}[0]{53~}

\begin{document}

\title{Calibrated Recommendations: Survey and Future Directions}

\author{Diego Corrêa da Silva}
\email{diego.correa@aau.at}
\orcid{0000-0001-7132-1977}
\affiliation{%
    \institution{University of Klagenfurt}
    \country{Austria}
}

\author{Dietmar Jannach}
\email{dietmar.jannach@aau.at}
\orcid{0000-0002-4698-8507}
\affiliation{%
    \institution{University of Klagenfurt}
    \country{Austria}
}

\renewcommand{\shortauthors}{da Silva and Jannach}

\begin{abstract}
The idea of calibrated recommendations is that the properties of the items that are suggested to users should match the distribution of their individual past preferences. Calibration techniques are therefore helpful to ensure that the recommendations provided to a user are not limited to a certain subset of the user's interests. Over the past few years, we have observed an increasing number of research works that use calibration for different purposes, including questions of diversity, biases, and fairness. In this work, we provide a survey on the recent developments in the area of calibrated recommendations. We both review existing technical approaches for calibration and provide an overview on empirical and analytical studies on the effectiveness of calibration for different use cases. Furthermore, we discuss limitations and common challenges when implementing calibration in practice.
\end{abstract}



\maketitle

\section{Introduction}
\label{sec:introduction}
Recommender systems are nowadays widely used on various online platforms to help users discover relevant content. Commonly, such systems make personalized recommendations based both on an individual user's past observed interests and based on preference patterns from a larger user community~\cite{Ricci:2010:RSH:1941884,jannachRSintroduction2011}. Historically, most published research on recommender systems from the last decades focuses on designing machine learning models that accurately place the assumedly most relevant items at the top of the recommendation list. However, already many years ago it was observed that ``\emph{being accurate is not enough}''~\cite{McNee:2006:Enough}, and that focusing solely on prediction or ranking accuracy may be actually detrimental for the user experience of a recommender system, for example because the resulting recommendations may be boring and lacking diversity or novelty~\cite{Kaminskas:2016:DSN:3028254.2926720}.

In this broader context of such ``beyond-accuracy''~\cite{GeDelgado-BattenfeldEtAl2010} considerations, Steck~\cite{Steck:2018:Calib} coined the term ``calibrated recommendations'' in a paper from 2018. He used the following intuitive example to illustrate this idea: ``\emph{When a user has watched, say, 70 romance movies and 30 action movies, then it is reasonable to expect the personalized list of recommended movies to be comprised of about 70\% romance and 30\% action movies as well.}'' The general idea of calibration in this example in the domain of movie recommendations is thus to match the distribution of movie genres in the set of recommendations with the distribution of the genres in the individual user's past preference profile. As a result, if a user had diverse preferences in the past, this diversity will be reflected in the recommendations after calibration, thereby avoiding monothematic recommendations. Notably, calibration is different from many previous approaches to increasing recommendation diversity like~\cite{Bradley-AICS-2001,ziegler2005diversity} in that the diversification process is personalized to the individual user.

Technically, calibration approaches are commonly implemented as \emph{reranking} techniques. This means that the outputs of any underlying recommendation algorithm are post-processed to ensure that the distributions of the item properties in the recommendation list and the user profile are aligned. The level of alignment of the distributions can be gauged with the help of measures from information theory like Kullback-Leibler divergence~\cite{Silva:2025:Benchmark}. Besides item features like movie genres, other properties of the items can be the target of the calibration process, for example an item's general popularity, which often is a relevant factor in the context of \emph{fairness} considerations of recommender systems~\cite{Abdollahpouri:2020:PopularityCalibration, Abdollahpouri:2021:UserCentered}.

However, while the idea of calibration as described by Steck is intuitive and useful for many use cases, it may also face challenges and limitations in certain situations. In real-world settings, for example, not all past preferences may be equally important, and older preferences may no longer be relevant and should thus not be considered for calibration. Also, at the level of individual users it may also happen that calibration actually \emph{reduces} diversity in case a user had a monothematic interest profile in the past. But there are also challenges from a practical perspective. Calibration by design creates a trade-off situation, where recommendation accuracy has to be balanced with a calibration target. Finding the right balance between these competing objectives can be challenging in practice. Also, when calibration is used to match a user's preferences in terms of item popularity it is a non-trivial process to determine appropriate thresholds values that are used to discriminate between different levels of popularity~\cite{Klimashevskaia:2023:Evaluating}.

Given these challenges, a variety of calibration-based approaches have been proposed over the past few years, and the impact of calibrated recommendations has been studied in a growing number of research works. With this present work, our goal is to provide a structured survey of these developments in an area of high practical relevance. To that purpose, we have systematically scanned the literature and identified \finalNbOfPapers relevant papers, which we discuss and categorize in this work. The paper is organized as follows. Next, in Section~\ref{sec:methodology}, we elaborate on our research methodology. We then provide an overview existing technical approaches for calibration in Section~\ref{sec:technical-approaches}. Afterwards, in Section~\ref{sec:evaluation}, we focus on the evaluation of calibration approaches and we review studies which have analyzed the effects of recommendations in offline and online settings. The paper continues with a discussion of existing solutions and open challenges in Section~\ref{sec:discussion} and ends with an outlook on future developments.

\section{Research Methodology and General Statistics}
\label{sec:methodology}

\paragraph{Identification of Relevant Papers}
We followed a systematic approach~\cite{Kitchenham:TechReport2007:SysReviews} to identify relevant research works. Specifically, we queried two bibliographic databases, namely the ACM Digital Library (DL) and Google Scholar with a pre-defined set of keywords. The ACM DL was chosen because many important publication venues for recommender systems research are related to ACM, e.g., the SIGIR, WWW, or RecSys conference series. Google Scholar was used to complement the search because its index is very comprehensive. The following query was used to identify papers that had relevant terms in the full text.

\begin{quote} \texttt{\small
(``calibrated recommendations'') OR (``calibrated recommenders'') OR (``calibrated recommendation'')  OR (``calibrated recommendation system'') OR (``calibrated recommender system'') OR (``calibrated recommender'') OR (``calibrated recommendation systems'') OR (``calibrated recommender systems'')}
\end{quote}

We limited the search to papers appearing since 2018, i.e., the year when Steck's paper was published. The search, when executed in May 2025, yielded 57 papers in the ACM DL and 356 papers from Google Scholar. In terms of inclusion criteria, we only retained works that build on Steck's notion of calibration, e.g., by proposing a new calibration method or by evaluating and analyzing the impact of calibrated recommendations. Correspondingly, we excluded works that only discuss calibration as related work or cite such works somewhere in the text, i.e., works without a specific contribution to calibration research. Furthermore, we did not consider preprints, e.g., from arxiv.org or ssrn.com, and Master's and PhD theses. Papers that were not written in English were also excluded. Papers published in conferences and later expanded in journals were considered as one paper.

Applying these inclusion and exclusion criteria left us with \initialNbOfPapers research papers. In addition, going through the references of these papers, we found two earlier works that propose ideas that are very similar to Steck's calibration approach, but which were not using the term calibration. These works were published in 2011 and 2017 by Oh et al.~\cite{Oh:2011:Novel} and by Jugovac et al.~\cite{Jugovac:2017:Efficient}, where the latter work is based on the first one. Oh et al.~propose a method called Personal Popularity Tendency Matching (PPTM), where the goal is to increase the novelty of the recommendations by aligning the popularity of the recommended items with the user's past popularity preferences. Like Steck, they propose a reranking procedure for this alignment. Differently from Steck, Oh et al.~use the \emph{Earth Mover's Distance} as a measure for miscalibration.\footnote{A very similar idea was independently developed later in~\cite{Abdollahpouri19Managing} to manage popularity bias through personalized reranking.} Jugovac et al.~then later propose an extension to this work in the context of music recommendations, in which multiple optimization goals can be considered in parallel. Adding the approaches presented in~\cite{Jugovac:2017:Efficient} and~\cite{Oh:2011:Novel} left us with \finalNbOfPapers papers that were finally considered in our analyses.

We note that calibration techniques are also related to early intent-aware recommender systems (IARS). IARS are designed to take time-varying user intents into account when making recommendations~\cite{jannach2024intent}. Early approaches in this area~\cite{Vargas2012Explicit,vargas_intent-oriented_2011} implemented intent-awareness (or: ``intent-orientation'') by considering the past interest of users in certain item features, which could for example be movie genres. To cover most of the possible user intents or interests, these approaches then work by diversifying the recommendations by ensuring that the recommendations cover many of the past interests, e.g., movie genres. Technically, this is commonly implemented by a re-ranking procedure that considers both the relevance of the items and a diversity component. A deeper discussion and analysis of intent-aware and calibrated recommendations can be found in~\cite{Kaya:2019:Intent}. While some intent-aware approaches are technically related to calibration, we did not include intent-aware models in our survey as their main target is typically to being able to cover short-term user interests.

\begin{figure}[h!]
    \centering
    \includegraphics[width=0.7\linewidth]{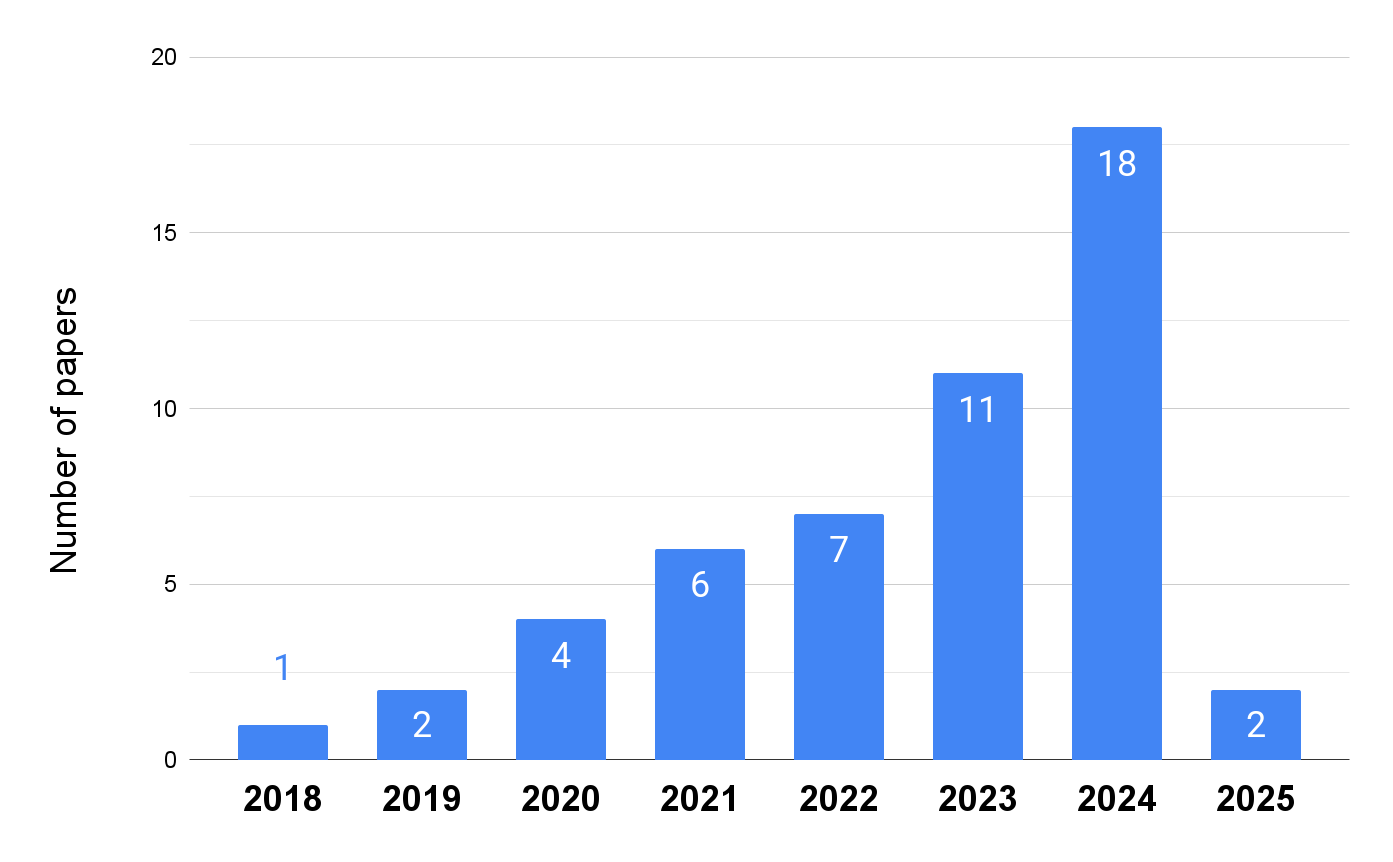}
    \caption{Number of publications per year on calibrated recommendations.}
    \Description[Single figure]{Figure shows eight bars, each bar is one year, and the number of publications in each year.}
    \label{fig:papersperyear}
\end{figure}

\paragraph{Statistics of Collected Papers}
We retrieved the metadata of the \finalNbOfPapers identified research works to analyze the number of publications per year, their publication venues, and the nature of their contributions. As shown in Figure \ref{fig:papersperyear}, the number of papers in this area has consistently increased over the past eight years, with the most significant growth occurring in 2024. This trend indicates that the field is experiencing strong annual growth.

\begin{figure*}[!ht]
	\centering
	\begin{subfigure}{0.48\textwidth}
		\centering
		\includegraphics[width=1\linewidth]{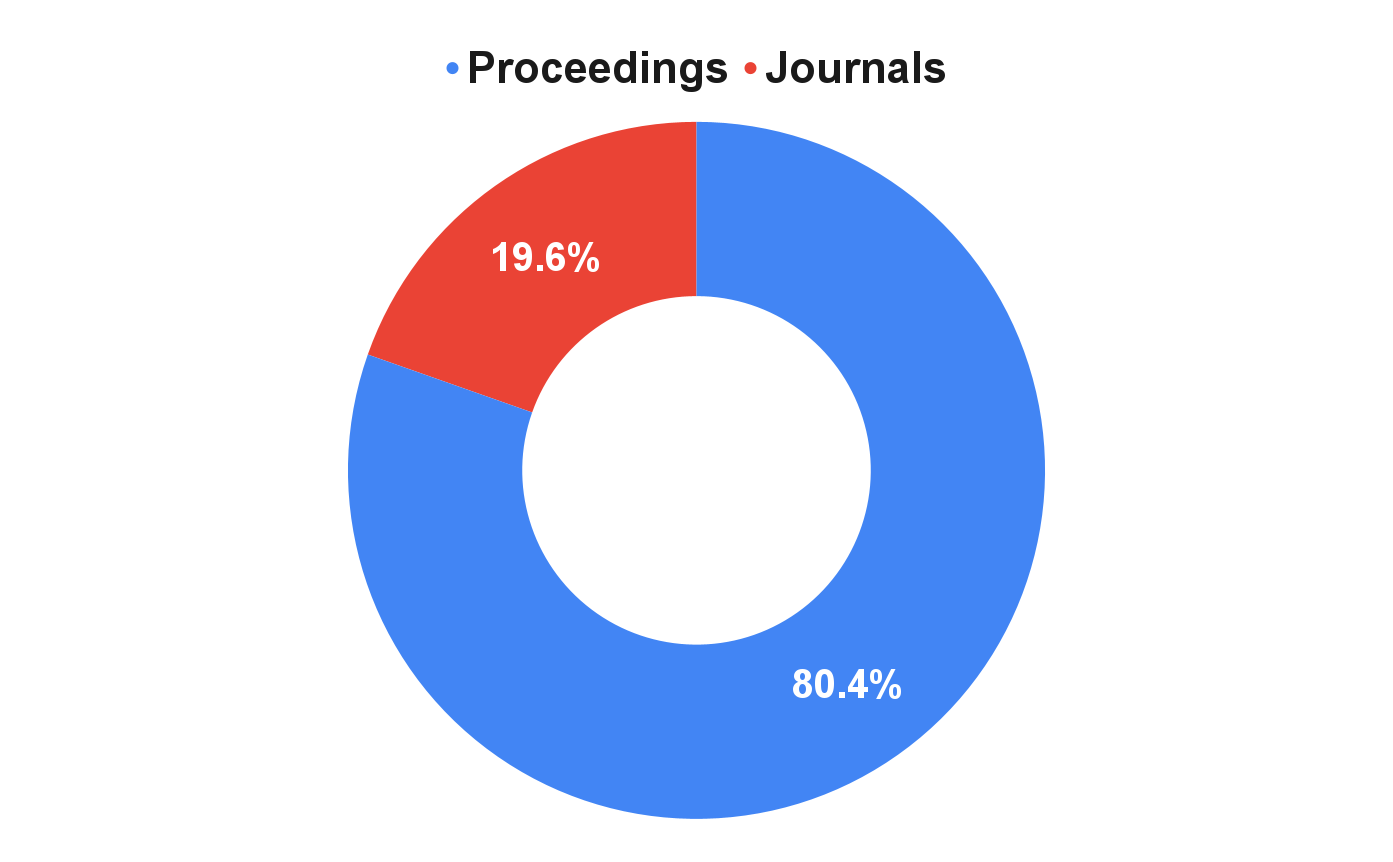}
		\caption{Proportion of proceedings and journals.}
		\label{fig:place}
	\end{subfigure}
	\begin{subfigure}{0.47\textwidth}
		\centering
		\includegraphics[width=1\linewidth]{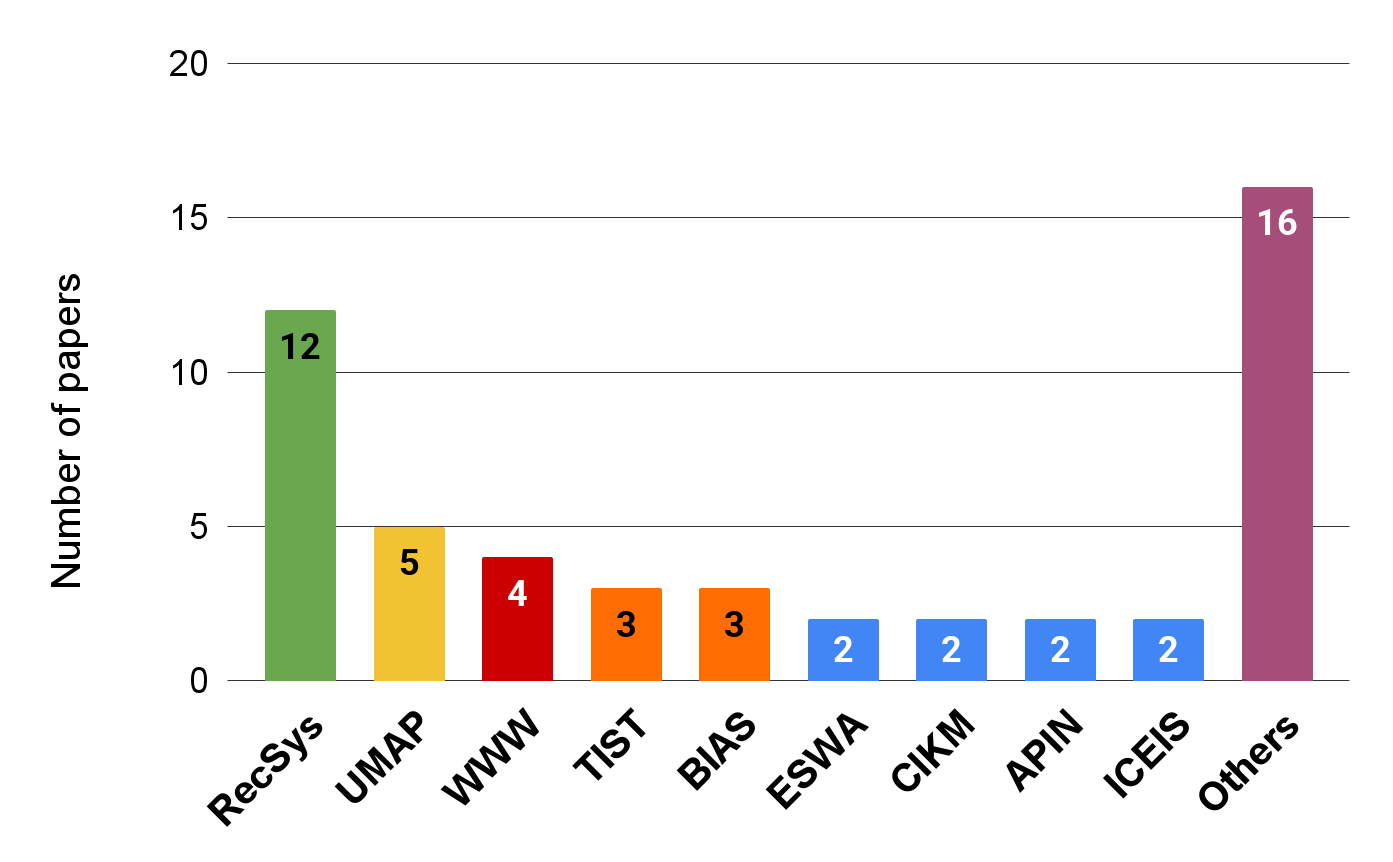}
		\caption{Venues publishing papers on calibrated recommendations.}
		\label{fig:venue}
	\end{subfigure}

	\caption{Publication venues.}
        \Description[Two figures]{Figure 1 shows the percentage of publications in Proceedings and Journal. Figure 2 shows eleven bars, where ten bars are the venues with the most number of publications and the last bar comprises all other venues.}
	\label{fig:publishingplaces}
\end{figure*}

Figure \ref{fig:place} illustrates the distribution of publications in proceedings and journals. As we can see, more than 80\% of the papers were published in conference or workshop proceedings and less than 20\% in journals. Most of publications are presented at major conferences, with a notable concentration at the ACM Conference on Recommender Systems (RecSys)\footnote{https://recsys.acm.org/} with 12 publications. Other prominent venues include ACM UMAP\footnote{https://dl.acm.org/conference/umap} and the ACM Web Conference\footnote{https://thewebconf.org/} conferences with five and four publications respectively. The journal with the most relevant publications is ACM Transactions on Intelligent Systems and Technology (TIST)\footnote{https://dl.acm.org/journal/tist}. Three papers appeared in the proceedings of a workshop on Advances in Bias and Fairness in Information Retrieval (BIAS)\footnote{https://link.springer.com/conference/bias}. The venues with two publications on calibration are Expert Systems with Applications (ESWA)\footnote{https://www.sciencedirect.com/journal/expert-systems-with-applications}, the ACM International Conference on Information \& Knowledge Management (CIKM)\footnote{https://dl.acm.org/conference/cikm}, Applied Intelligence (APIN)\footnote{https://link.springer.com/journal/10489} and the International Conference on Enterprise Information Systems (ICEIS)\footnote{https://link.springer.com/conference/iceis}. The ``Others'' category in Figure \ref{fig:venue} represents all venues with only one paper published on calibrated recommendations.

\begin{figure}[h!]
    \centering
    \includegraphics[width=0.7\linewidth]{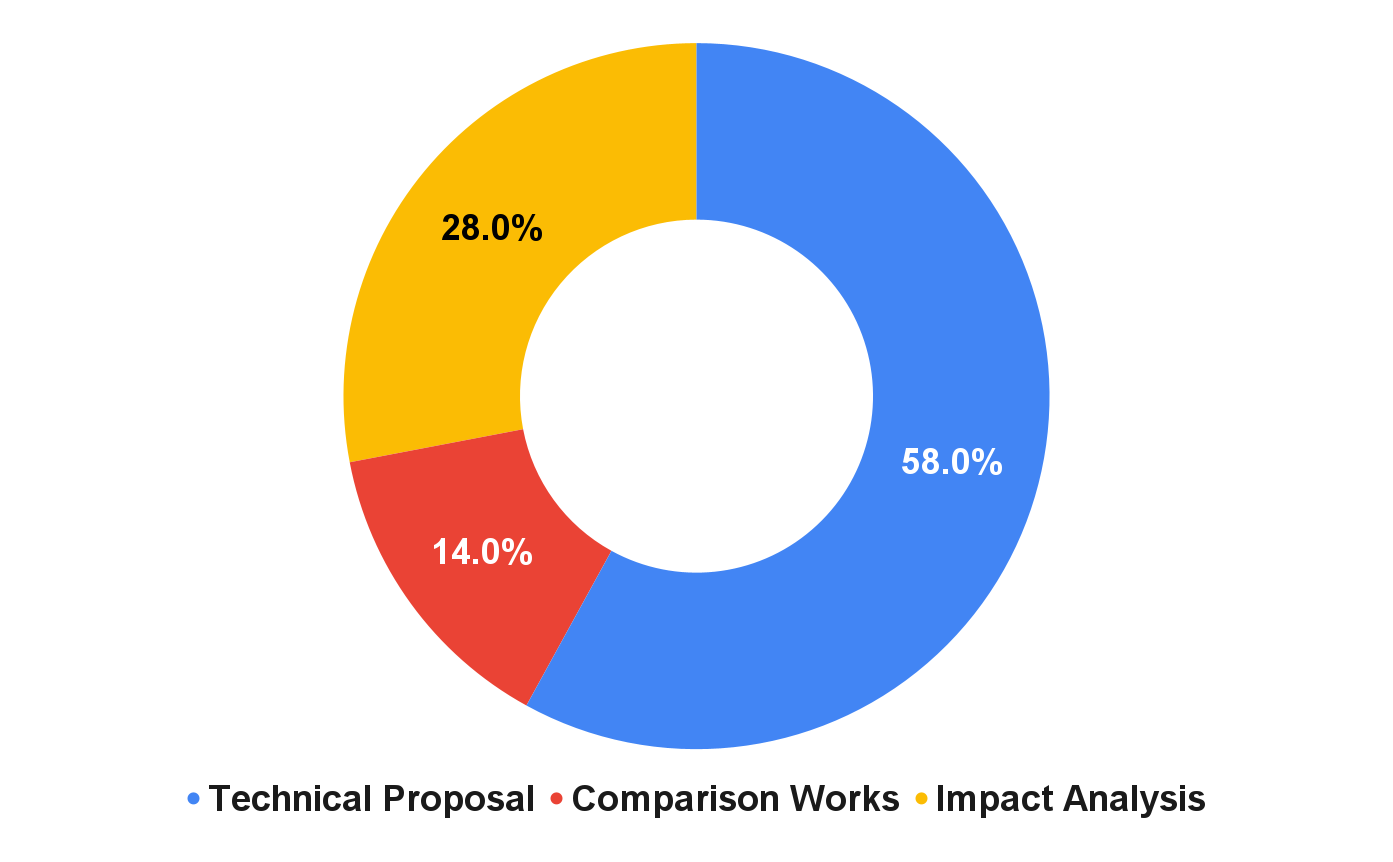}
    \caption{Proportion of papers classified based on the type of contribution.}
    \Description[Single figure]{This figure shows the highest emphasis (58.0\%) placed on "Impact Analysis,". The "Technical Proposal" follows at 28.0\%, while "Comparison Works" holds the least weight at 14.0\%.}
    \label{fig:approach}
\end{figure}

Among the \finalNbOfPapers research papers, we identified recurring patterns that allowed us to classify them into three distinct categories (Figure \ref{fig:approach}): technical proposals, impact analyses, and comparison works. The focus of each category is detailed below, and all relevant studies will be discussed in the following sections.

\begin{itemize}

    \item{\textbf{Technical Proposal}: Papers in this category introduce new technical contributions to the field, including modifications or improvements to existing calibration methods. The majority of the analyzed papers fall into this category, highlighting the research community's focus on expanding and refining calibration techniques.}

    \item{\textbf{Impact Analysis}: These studies examine specific aspects of calibrated recommendations, such as end user perceptions, algorithm behavior and biases (by popularity, stereotypes, human gender, age, and country), or system performance, without proposing new calibration methods. This category represents the second-largest share of publications, indicating a strong interest in understanding how calibration or its absence impact the system's recommendations and the user experience.}

    \item{\textbf{Comparison Works}: This category includes studies that compare different calibrated recommendation methods, evaluating their strengths and weaknesses. These comparisons often assess calibration effectiveness with and without additional post-processing techniques. It is the least common type of publication, suggesting that while new calibration methods are frequently introduced, fewer studies focus on systematically comparing their effects.}

\end{itemize}

\section{Technical Approaches to Calibration}
\label{sec:technical-approaches}
In this section, we outline the general architecture of most calibration approaches, using Steck's proposal as a foundational reference. We then also summarize the various implementations of individual components within this architecture, relating to works with proposals for the component. This technical section formally develops the calibrated recommendation framework, with Table \ref{tab:symbols} summarizing the notation used throughout.

\begin{table}[t!]
    \centering
    \caption{Formal notation used in the paper.}
    \label{tab:symbols}
    \begin{tabular}{l|l}
        \hline
        \textbf{Symbol}  & \textbf{Description}                                                                                                                                                       \\ \hline
        $u \in U$        & $u$ represents a unique user; $U$ is the set of all users. \\
        $i \in I$        & $i$ represents a unique item; $I$ is the set of all items. \\
        $g \in G$        & $g$ represents a unique genre; $G$ is the set of all genres. \\
        $H_{u}$          & The preference profile of $u$ derived from the user's historical interactions with items.\\
        $L_{u}$          & A list for recommendations for $u$. \\
        $L^{*}_{u}$      & A calibrated list of recommendation for $u$. \\
        $N$              & Size of the list of calibrated recommendations. \\
        $n$              & Temporary calibrated recommendation list size. \\
        $w_{u, i}$       & The relevance value (e.g., a rating) given by $u$ to item $i$. \\
        $\hat{w}_{u, i}$ & The predicted relevance value by the recommendation algorithm. \\
        $\lambda$        & The trade-off weight value. \\
        $CL(\cdot\ ,\cdot)$             & The calibration function used to measure the distance between two distributions. \\
        $REL(\cdot)$     & A function that returns the relevance of a given list of items. \\
        $P(\cdot)$       & A function to derive the \emph{target} distribution from the user profile.\\
        $Q(\cdot)$       & A function to derive a \emph{realized} distribution from a recommendation list. \\
        $P_{u}$          & The target distribution for user $u$. \\
        $Q_{u}$          & The realized distribution for user $u$. \\
        $p(g|u)$         & The distribution value of genre $g$ in $u$'s profile. \\
        $q(g|u)$         & Represents the distribution value of genre $g$ in $u$'s recommendation list. \\ \hline
    \end{tabular}
\end{table}

\subsection{A General Architecture for Calibration Approaches}
\label{sec:basic_arch}
Calibration is commonly implemented as a post-processing approach that operates independently of an underlying base recommender algorithm. Given a ranked list of candidate items, the calibration process consists of re-ranking the list to achieve a given calibration goal. Most studies identified in our survey follow Steck's framework, which can be decomposed into several key steps for which different choices can be made~\cite{Silva:2023:Protocol}: \emph{Baseline Ranking}, \emph{Candidate Selection}, \emph{Measuring Aggregate Relevance}, \emph{Determining Target and Realized Distributions}, \emph{Choosing the Calibration Measure}, \emph{Trade-off Balance Modeling}, and \emph{Optimized Item Selection}.

\subsubsection{Baseline Ranking}
In the \emph{Baseline Ranking} step, any algorithm can be applied to determine a personalized item ranking for a given user $u$. Commonly, collaborative filtering techniques, e.g., based on matrix factorization, are applied that solely work on a given user-item interaction matrix. However, content-based or hybrid techniques can also be used, as long as these techniques provide a personalized ranking using some sort of relevance score. Notably, different baseline ranking techniques may lead to different levels of initial miscalibration of the recommendations, e.g., in case when one baseline techniques generates more diverse recommendations `by design' than another one.

\subsubsection{Candidate Selection}
The next step in the process is \emph{Candidate Selection}, i.e., the selection of items that should be considered in the subsequent calibration step. In all studied papers, this process simply consists of considering the first \emph{k} elements recommended by the baseline ranker for each user. The most common number of items in the candidate set is 100. However, there are also works that use smaller candidate item sets (e.g., 10 or 20 in~\cite{Jugovac:2017:Efficient}) or larger ones (e.g., 1~000 in~\cite{Seymen:2021:Constraint, Gomez:20024:MOREGIN}). Intuitively, when using a smaller candidate set from a relevance-ordered list, the potential compromise on relevance can be limited. At the same time, however, the options for calibrating the list are reduced as well.

\subsubsection{Measuring Aggregate Relevance}
The core of calibration is to balance the relevance of a set of recommended items and its alignment with past preference patterns. In a next step, we therefore need to quantify the relevance of a given \emph{set} of recommendations $L_{u}$ through a relevance function $\text{REL}(L_{u})$. A basic and most frequently implemented approach is to simply rely on the relevance scores computed by the baseline ranker and use the sum of these scores as an aggregate relevance measure, more formally:

\begin{equation}
    \text{REL}(L_{u}) = \sum_{i \in L_{u}} \hat{w}_{u, i}.
    \label{eq:relevance_comp}
\end{equation}

In the equation, the relevance score for a given user $u$ and an item $i$ is denoted as $\hat{w}_{u, i}$. Depending on the use case, the relevance scores can be predicted user ratings, e.g., on a scale from 1-5, or other forms of relevance predictions, e.g., estimated probabilities that a user will do a certain action on an item.

Steck argues that alternative metrics such as Mean Reciprocal Rank (MRR) and Normalized Discounted Cumulative Gain (NDCG) can be used to assess recommendation relevance. While most studies rely on Steck's original proposal and implement Equation \ref{eq:relevance_comp}, some researchers have explored modifications to enhance the system. Sacharidis et al.~\cite{Dimitris:2019:Common}, for example, introduce an alternative utility function inspired by the NDCG, aiming to improve ranking quality during the calibration step. da Silva and Durão~\cite{Silva:2025:Benchmark} conducted a benchmark study to systematically compare the widely adopted summation-based approach with the NDCG, and their results indicate that both methods exhibit similar performance.

An alternative approach by Zhao et al.~\cite{Zhao:2020:Distortion} shifts the focus to recommending users for items. In this work, the authors modify Steck's relevance function to better align with the task of recommending the most suitable users for each item. The relevance value for an item-user pair is not computed directly as in Steck's proposal, but rather inferred indirectly from the user's ranking position for the item. Expanding the previous idea, Zhao et al.~\cite{Zhao:2021:Rabbit} introduce a new relevance function in which an item's relevance decreases based on its position in the recommendation list. In this follow-up study, they propose TecRec, a method that incorporates calibration into a sequential recommendation setting. Further refinements have been made in the context of sequential recommendation also in other works. For example, Chen et al.~\cite{Chen:2022:DACSR, Jiayi:2023:LongTail} incorporate the cross-entropy loss function as a relevance measure, although without detailed justification. Finally, in the context of top-n recommendation problems, Abdollahpouri et al.~\cite{Abdollahpouri:2023:MinCost} formulate the calibrated recommendation task as a graph model, where the relevance scores are the edge weights. These relevance scores are incorporated into a maximum weight assignment problem, where the objective is to select a set of items for a recommendation list that maximizes the total relevance.

\subsubsection{Determining Target and Realized Distributions}
We recall that the goal of calibration is to ensure that the made recommendations match the user's past preference patterns in terms of their distribution of certain item features. As such, the \emph{item features of interest} must be defined in a first step. Various calibration approaches are evaluated in the movie domain, and the movie \emph{genre} is thus used as a calibration target. Other works consider item categories instead~\cite{Nazari:2022:Podcast}. Furthermore, ``meta-level'' item attributes, such as their popularity were considered as well~\cite{Abdollahpouri:2021:UserCentered, Klimashevskaia:2022:Mitigating, Klimashevskaia:2023:Evaluating, Lesota:2022:Cross_Group, Lesota:2023:Perceived, Ungruh:2024:Mittigation}. Typically, exactly one item feature is used for calibration. An exception is the work by Jugovac et al.~\cite{Jugovac:2017:Efficient}, who present an approach in which user preferences along multiple dimensions can be considered in parallel.

To match the distributions of the user preferences, e.g., in terms of the item \textit{genre}, and the recommendations, a procedure has to be defined how these distributions are derived. In the case of movie genres, which we use for illustration purposes here, we note that a movie can belong to multiple genres. In such cases of multi-valued item features, the \emph{proportion} of a movie belonging to a certain genre is usually computed first. The most commonly implemented approach is to simply divide $1$ by the number of genres that are associated with a given item. For instance, the proportion of a specific genre $g$ of an item $i$ with four genres is calculated as $\frac{1}{4} = 0.25$. In this approach, each genre assigned to an item has an equal weight. Let us denote the set of genres assigned to an item $i$ with $g(i)$, and $|g(i)|$ the size of this set. We can then define the proportion of a genre for a given item as follows:

    \begin{equation}
        \text{$prop(g|i)$} = \frac{1}{|g(i)|}.
        \label{eq:genre_proportionally}
    \end{equation}

\paragraph{Target Distribution}
The \textit{Target Distribution}, denoted as $P(H_{u})$ following~\cite{Steck:2018:Calib}, represents the ideal distribution of item attributes (e.g., genres) based on the user's past interactions $H_u$. It captures the proportions in which different genres should appear in the recommendation list to best match the user's preferences. To compute $P(H_{u})$, we determine for each of the existing genres in the dataset, denoted as $G$, a value $p(g|u)$ that represents the ``importance'' of this genre for this user. Specifically, $p(g|u)$ is computed by considering all items in the user's past preference profile, and it both takes into account the genre proportion $prop(g|i)$ of each item and the user's preference towards this item, e.g., the user's explicit item rating. Overall, the derivation of the target distribution can be formally described as follows\footnote{The denominator in the equation is used for normalization.}:

    \begin{equation}
        \text{P($H_{u}$)} = (\forall g \in G)\ p(g|u) = \frac{\sum_{i}^{H_{u}}\mathbb{1}(g \in g(i))w_{u,i}\cdot prop(g|i)}{\sum_{i}^{H_{u}}\mathbb{1}(g \in g(i))w_{u, i}}.
        \label{eq:target_dist}
    \end{equation}

\paragraph{Realized Distribution}
The \textit{Realized Distribution}, denoted as $Q(L_{u})$, represents the actual distribution of item attributes in the generated recommendation list $L_{u}$ for a user $u$. In our example, it is again computed based on the genres of the recommended items. The subsequent goal of the calibration technique is to narrow the gap between the realized and the target distributions, e.g., by re-ranking the recommendation list. Technically, the computation is identical to the one for the target distribution in Equation~\ref{eq:target_dist}, with the difference that we are now considering the user-specific importance of each genre of an item in the recommendation list $q(g|u)$ using the user's predicted rating $\hat{w}_{u, i}$ for each item:

    \begin{equation}
        \text{Q($L_{u}$)} = (\forall g \in G)\ q(g|u) = \frac{\sum_{i}^{L_{u}}\mathbb{1}(g \in g(i))\hat{w}_{u,i}\cdot prop(g|i)}{\sum_{i}^{L_{u}}\mathbb{1}(g \in g(i))\hat{w}_{u,i}}.
        \label{eq:realized_dist}
    \end{equation}

We note that using this approach to compute $Q(L_{u})$ may result in a tricky situation during calibration in cases where a certain genre is very infrequent in the user profile, and thus the target distribution. In such cases, the genre might not be considered at all during the calibration process, because adding an item of this genre might immediately lead to an increase in the level of miscalibration, which is caused by overrepresenting this genre in the realized distribution. To mitigate this issue, Steck proposes a smoothing re-balancing approach $\tilde{Q}(L_{u})$ that ensures that no genre has a value of zero in the realized distribution.

    \begin{equation}
        \tilde{Q}(L_{u}) = (\forall g \in G)\ \tilde{q}(g|u) = (1-\alpha)\cdot q(g|u) + \alpha \cdot p(g|u).
        \label{eq:realized_dist_alt}
    \end{equation}

A commonly used value for $\alpha$ is 0.01, which prevents extreme sparsity in the distribution while maintaining the overall proportional structure.

\paragraph{Alternatives to Modeling User Distributions} The core concept of calibration relies on the derivation of distributions, making the design of an effective distribution a critical research focus. Thus, several works propose alternative distribution formulations. Jeon et al.~\cite{Jeon:2024:LeapRec} introduce an additional hyperparameter weighting factor to refine the distribution derivation process for sequential recommendation contexts. Specifically, the weighting factor can be used to give more emphasis to more recent user interactions. Jiayi et al.~\cite{Jiayi:2023:LongTail} simplify Steck's approach by deriving distributions based on item popularity, distinguishing between head and long-tail items. Chen et al.~\cite{Chen:2022:DACSR} integrate the distribution derivation process directly into the recommender algorithm, applying a softmax function to the target distribution. Differently from these personalized approaches, Abdollahpouri et al.~\cite{Abdollahpouri:2021:Target} propose to use a system-wide target distribution determined by domain experts, which is combined with the target distribution to balance both system and user interests. Zhao et al.~\cite{Zhao:2021:Rabbit} apply Steck's method but incorporate a Long Short-Term Memory (LSTM) module to predict the future user preference distribution based on the evolution of the historical distribution. Finally, da Silva and Durão~\cite{Silva:2025:Benchmark} demonstrate that a simple normalization step, ensuring a properly normalized distribution, consistently improves both precision and calibration compared to Steck's proposal. Later on, da Silva et al.~\cite{Silva:2025:TimeEntropy} incorporate entropy and temporal factors into the distribution derivation process, finding that entropy-based distributions can outperform even the normalized version.

\subsubsection{Choosing the Calibration Measure}
\label{sec:calibration_measure}
Given the goal of creating a recommendation list that leads to a realized distribution $Q$ that is close to the target distribution $P$, i.e., $Q \approx P$, a measure is needed to quantify the discrepancy between $P$ and $Q$. Various measures exist in the literature to quantify the distance between distributions~\cite{Cha:2007:survey}. In his proposal, Steck relied on the Kullback-Leibler (KL) divergence, as this measure possesses three useful properties for calibration~\cite{Steck:2018:Calib}: (a) it is zero when $P=Q$; (b) it is sensitive to small differences between the distributions and (c) it favors uniform and less extreme distributions.

Given a target distribution $P$ and a realized distribution $Q$, the KL divergence is defined as follows.

    \begin{equation}
        \text{CL}_{KL}(P_{u}, Q_{u}) = \sum_{} p(g|u)\cdot log\frac{p(g|u)}{q(g|u)}.
        \label{eq:calibration_comp}
    \end{equation}

Most studies follow Steck's approach and implement the Kullback-Leibler (KL) divergence as the primary calibration measure. Another widely adopted measure is the Jensen-Shannon (JS) divergence, which has been used in several works~\cite{Abdollahpouri:2021:UserCentered, Abdollahpouri:2021:Target, Naghiaei:2022:Towards, Klimashevskaia:2022:Mitigating, Lesota:2022:Cross_Group, Nazari:2022:Podcast, Vrijenhoek:2022:RADio, Lesota:2023:Perceived, Klimashevskaia:2023:Evaluating, Ungruh:2024:Mittigation, Silva:2025:Benchmark}.

Beyond these commonly used approaches, some studies propose alternative distance measures. In an early work, Oh et al.~\cite{Oh:2011:Novel} proposed to use the Earth-Mover's Distance (EMD) as a way to assess the distance between two distributions, where the EMD measure corresponds to the amount of `work' needed to transform one distribution into the other. Sacharidis et al.~\cite{Dimitris:2019:Common} later introduce a measure called Normalized Fairness, which is the divergence between $P$ and $Q$ normalized by the highest possible upper bound. The rationale of the approach is however not further detailed in the paper. Seymen et al.~\cite{Seymen:2021:Constraint} propose to use the Weighted Total Variation of $P$ and $Q$ due to the non-linearity of the KL-divergence measure. Chen et al.~\cite{Chen:2023:Triple} integrate calibration directly into a neighbor-based recommender algorithm and develop an approach that applies the KL-divergence measure across three different ``orders'' of calibration. The first-order distance measures the target distribution distance between two users. The second order extends this by evaluating the target distribution distances among three users, using one of these users as a reference point to obtain the distance value. The third order measures the distance between the average genre ratings for two users. By combining these three orders, the approach enhances the neighbor selection process, ultimately improving the quality of the recommendations.

In principle, any distance measure can be used in this component. Following this idea, da Silva and Durão~\cite{Silva:2025:Benchmark} conduct a benchmark study comparing 57 distance measures in a movie recommendation context. Their findings reveal that the same four measures, Vicis Emanon2, Kulczynski S, Kumar Johnson, and Additive Chi, consistently achieve the best performance across different datasets. This suggests that alternatives to KL and JS divergence may yield superior results.

\subsubsection{Trade-off Balance Modeling}
Calibration commonly involves a trade-off situation, where the aggregated relevance of the recommended items competes with the goal of covering past user preferences well. This trade-off must be modeled accordingly, using a formalism that quantifies the overall quality or utility of a (calibrated) recommendation list by taking into account both factors. The most common formulation in the literature to obtain an optimal calibrated list $L^{*}_{u}$ is as follows:

    \begin{equation}
        L^{*}_{u} = \arg\max_{L, |L|=N} (1 - \lambda)\cdot \text{REL}(L_{u}) - \lambda \cdot\text{CL}(P_{u}, Q(L_{u})).
        \label{eq:utility}
    \end{equation}

Equation \ref{eq:utility} consolidates all functions into a single utility value, which serves as the foundation for evaluating the quality of the calibrated recommendation list. While many studies adhere to this formulation, several alternative trade-off implementations have been proposed. Chen et al.~\cite{Chen:2022:DACSR} extend the original trade-off function by incorporating additional relevance and distribution values using an algorithm-dependent technique. A detailed explanation of the intuition behind the specific approach is however not given. Other variations include the work by Naghiaei et al.~\cite{Naghiaei:2022:Towards}, who adopt a Mixed-Integer Linear Programming (MILP) approach to optimize the trade-off function, leading to a slight performance improvement. da Silva and Durão~\cite{Silva:2023:Protocol} introduce a logarithmic trade-off function that modifies Steck's original proposal by incorporating a logarithmic transformation and personalized bias scores. The authors argue that using a logarithmic function helps to mitigate large discrepancies in the calibration level. Additionally, by incorporating user bias scores, the method prioritizes the selection of items whose predicted ratings are close to the user's average rating, thereby reducing discrepancies between user preferences and the recommended list.

Seymen et al.~\cite{Seymen:2021:Constraint}, in contrast, modify the trade-off calculation by computing a global trade-off value across all users, rather than applying it individually. This approach aims to improve the overall diversity of the recommended items. Instead of creating independent recommendation lists for each user, the method inserts one item into each user's list and then evaluates the system-wide level of miscalibration. This process is repeated incrementally—adding another new item to all lists and re-evaluating—allowing the system to progressively reduce miscalibration at a global level.

Coming back to the most frequently used trade-off balance approach as described in Equation \ref{eq:utility}, the trade-off weight $\lambda$ is typically a constant value that is used for all users. A weight of $0.0$ focuses solely on accuracy, effectively bypassing calibration. Conversely, a weight of $1.0$ maximizes calibration, constructing the most calibrated list possible. A value of $0.5$ considers both factors equally. Many studies evaluate this weight using varying values from $0.0$ to $1.0$ in increments of $0.1$, aiming to identify an optimal balance of calibration and accuracy.

While most studies treat $\lambda$ as a constant value that is predefined by domain experts, some works have explored personalized approaches to determine this trade-off weight dynamically. Wang et al.~\cite{Wang:2021:Deconfounded} integrate calibration into the recommender algorithm as part of a deconfounding approach to mitigate bias amplification. Their method personalizes $\lambda$ by normalizing divergence values, using the minimum and maximum divergence values across all users. Lesota et al.~\cite{Lesota:2022:Cross_Group} take a different approach by dynamically adjusting $\lambda$ throughout the calibration step, seeking a balance between calibration and relevance that recomputes the $\lambda$ value as items are added to the list. da Silva et al.~\cite{Silva:2021:Exploiting} propose two methods for deriving a user-specific value for $\lambda$, arguing that individual users have different preferences regarding the degree of calibration in their recommendations. Different studies show~\cite{Silva:2021:Exploiting, Silva:2023:Protocol, Silva:2025:Benchmark} that using personalized weights can lead to improvements, depending on the recommender algorithm.

\subsubsection{Optimized Item Selection}
Determining the optimal list $L^{*}_{u}$ is an NP-hard combinatorial optimization problem, which does not scale well given the large item catalogs in many application settings. Usually, the search for an optimal (or sufficiently good) solution is thus limited to a smaller candidate set, as described above, e.g., the top-100 most relevant items determined by the baseline recommendation technique.

Technically, often greedy re-ranking techniques are applied to speed up the search for good-enough solutions. Steck proposes to use the Surrogate Submodular Greedy algorithm, which achieves a $(1 - 1/e)$ optimality guarantee\footnote{The symbol $e$ is Euler's number.} for optimal item selection. The algorithm starts with an empty list and iteratively adds items at each list position by selecting the next item that maximizes the utility of the list according to Equation~\ref{eq:utility}. This process is continued until a list of the desired length is obtained. The algorithm's complexity increases with the number of candidate items and the size of the requested recommendation list, making it computationally expensive, especially when dealing with large item catalogs.

Several alternative algorithms have been proposed to improve accuracy, calibration, and computational efficiency. For instance, the Top-Z algorithm implemented in~\cite{Zhao:2020:Distortion, Zhao:2021:Rabbit} offers a smooth improvement over the Surrogate Submodular algorithm. Top-Z introduces two additional parameters: the number of unique genres in the system ($|G|$) and a constant $Z$ defined by the system specialist. The product of these two parameters determines the number of candidate items to be re-ranked, effectively controlling the re-ranking pool size and improving the balance between computational efficiency and calibration performance as reported in paper results. Starychfojtu and Peska~\cite{Starychfojtu:2020:SmartRecepies} introduce a fuzzy logic-based approach by implementing the Fuzzy D'Hondt algorithm for calibration re-ranking. They argue that the Fuzzy D'Hondt algorithm not only ensures a calibrated selection but also provides a fair ordering of the recommended items. Additionally, some studies frame item selection as a branch-and-bound problem~\cite{Seymen:2021:Constraint, Naghiaei:2022:Towards}. In particular, Seymen et al.~\cite{Seymen:2021:Constraint} adopt this approach in their method, which fills one position across all users' recommendation lists simultaneously. As the algorithm inserts an item into a user's list, it immediately updates the system-wide miscalibration value before proceeding to the next insertion. This process aligns with the branch-and-bound workflow, enabling dynamic recalculations of the miscalibration across all lists during list construction. Naghiaei et al.~\cite{Naghiaei:2022:Towards} rely on the use of a branch-and-bound approach as a consequence of formulating the problem as a Mixed-Integer Linear Programming (MILP) task. Abdollahpouri et al.~\cite{Abdollahpouri:2023:MinCost} observe that MILP cannot guarantee an optimal solution efficiently in practice. To address this limitation, they argue for the need to adopt a more computationally efficient algorithm. Based on this motivation, the authors reformulate the calibration recommendation problem as a maximum flow optimization task. They introduce a new method called Minimum-Cost Flow (MCF), which models the recommendation process using a graph structure, where costs are assigned to the edges between adjacent nodes. The objective of MCF is to determine the least-cost path through the graph that yields a well-calibrated recommendation list. In a subsequent study, Naghiaei et al.~\cite{Naghiaei:2024:BeyondAccuracy} also apply MILP and branch-and-bound techniques, aiming to assess the improvements of their proposal across a broader set of metrics beyond accuracy. Finally, Kleinberg et al.~\cite{Kleinberg:2024:Decaying} propose a novel algorithm for generating an optimal calibrated recommendation list, though their work remains theoretical and lacks empirical validation.

Overall, four major algorithmic techniques have been investigated in calibrated recommendation: greedy algorithms, linear programming, branch-and-bound methods, and fuzzy logic. These approaches highlight the flexibility in solving the item selection problem, although trade-offs exist between computational cost and the effectiveness of accuracy and calibration. Jeon et al.~\cite{Jeon:2024:LeapRec} present a time complexity analysis of various proposals, including those from\cite{Steck:2018:Calib, Abdollahpouri:2023:MinCost, Seymen:2021:Constraint}, alongside their own approach, which is also based on a greedy strategy. They show that the MCF method introduced in~\cite{Abdollahpouri:2023:MinCost} exhibits the highest time complexity, making it the slowest among the compared methods. The MILP-based approach proposed in~\cite{Seymen:2021:Constraint} follows as the second slowest. In contrast, Steck's original approach~\cite{Steck:2018:Calib} and Jeon et al.'s approach~\cite{Jeon:2024:LeapRec} demonstrate similar and significantly better time complexities. These findings are further supported by results reported in~\cite{Mo:2024:Graph}.

\subsection{Alternative Implementation Approaches}
While the majority of the studies in the literature rely on the general architecture described in the previous section, a number of alternative approaches can be found as well. Specifically, we have identified a number of works that support the parallel consideration of multiple optimization objectives. Furthermore, a few works can be found that do not rely on a post-processing (re-ranking) approach for calibration.

\subsubsection{Supporting Multiple Optimization Objectives}
Most calibration studies use a single categorical attribute---such as the item's genre---as the basis for deriving the target distribution and generating a recommendation list that aligns with it. However, some studies go further by incorporating multiple objectives simultaneously. 

In a work preceding Steck's proposal, Jugovac et al.~\cite{Jugovac:2017:Efficient} propose a parameterizable re-ranking technique to balance multiple objectives like accuracy, diversity, and item popularity. In their approach, the level of miscalibration is individually measured for each of these objectives. During the calibration process, the effects of re-ranking an item are evaluated for each of these dimensions, and a potential re-ranking move is only applied in case it leads to an aggregated improvement when considering all dimensions.

In a more recent work, Gomez et al.~\cite{Gomez:20024:MOREGIN} propose a system that simultaneously optimizes accuracy, provider fairness, and calibration. Their approach introduces adjustments at both the global level focused on provider fairness, and the individual level focused on calibration. The core intuition is that the post-processing phase should ensure that the recommendations satisfy all three objectives simultaneously. Similarly, Treuillier et al.~\cite{Treuillier:2024:ADF} introduce a framework aimed at simultaneously optimizing accuracy, diversity, and calibration, ensuring that enhancements in diversity and calibration do not compromise recommendation accuracy. They argue that diversity and calibration are as important as accuracy and should not be treated as secondary objectives.

Sacilotti et al.~\cite{Sacilotti:2023:Popularity} and Souza et al.~\cite{Souza:2024:Popularity} simultaneously investigate whether users prioritize genre calibration or popularity calibration and personalize the re-ranking process accordingly. Their approach adapts the recommendation list based on individual user preferences regarding these two calibration factors. In another study, Souza et al.~\cite{Souza:2024:TwoStage} introduce a two-stage post-processing approach, where the list is first re-ranked based on popularity calibration, followed by a second re-ranking step that applies genre calibration. This sequential method aims to control both factors more effectively while maintaining recommendation quality.
Finally, Naghiaei et al.~\cite{Naghiaei:2024:BeyondAccuracy} extend multi-objective recommendation further by optimizing accuracy, calibration, novelty, surprise, coverage, and redundancy simultaneously. They define redundancy as a measure of how many redundant categories a user has interacted with. Their approach integrates these objectives to generate a well-balanced recommendation list that satisfies diverse user and system constraints.

\subsubsection{Alternatives to Re-ranking}
Most calibration proposals rely on a post-processing step. However, some approaches integrate calibration directly into the recommender algorithm, eliminating the need for post-processing. These studies highlight the benefits of embedding calibration within different recommendation paradigms, enhancing both efficiency and effectiveness while avoiding the computational overhead of a separate post-processing step.

Chen et al.~\cite{Chen:2022:DACSR} present the Decoupled-Aggregated Calibrated Sequential Recommendation (DACSR) method, which incorporates a divergence-based calibration measure directly into the loss function. To generate calibrated recommendation lists, the authors design a loss function that enforces consistency between the recommended sequence and the user's historical interaction sequence, ensuring that the calibration objective is considered during model training. In another work, Chen et al.~\cite{Jiayi:2023:LongTail} explore the intersection of calibration, long-tail popularity, and session-based recommendation. The authors use two categories to derive the distributions: tail items, representing less popular items, and head items, representing the most popular ones. The core idea is to balance these two categories according to the user's session provided as input.

Jeon et al.~\cite{Jeon:2024:LeapRec} also focus on sequential problems and introduce a novel method designed to improve recommendation quality while maintaining efficiency. Their approach consists of two main steps. The first one is backbone integration, where calibration is embedded within the recommender algorithm to align candidate items with the user's preference distribution while preserving accuracy. The second one is re-ranking for relevance, where the recommendation list is re-ranked to prioritize relevance without compromising calibration, seeking to include relevant item in the top and calibration in the bottom of the recommended list.

The integration of calibration into a neighborhood-based algorithm is proposed in~\cite{Chen:2023:Triple}, where the calibration trade-off is considered during neighbor selection. This approach utilizes the calibration distance measure to consider genre distribution differences between the target user and their nearest neighbors, as well as the deviation in average rating behavior.

A graph-based approach is presented by Mo et al.~\cite{Mo:2024:Graph}, who develop a calibrated recommendation system for Point of Interest (POI) suggestions. Their approach integrates calibration into a graph convolutional network (GCN), eliminating the need for post-processing by directly generating calibrated recommendations through the model itself. Wang et al.~\cite{Wang:2021:Deconfounded}, finally, compare calibration with four types of post-processing approaches and introduce the Deconfounded Recommender System (DecRS) model. This model is designed to extend the calibration idea to alleviate bias amplification directly within the recommender algorithm. DecRS leverages a causal graph framework, employing backdoor adjustment to mitigate the influence of confounding factors and to thereby improve the quality of the recommendations.

\subsubsection{Recommending Users For Items}
Most calibrated recommendation systems follow the traditional approach of recommending items to users. In contrast, Zhao et al.~\cite{Zhao:2020:Distortion}, as discussed earlier, reverse this process by focusing on user-to-item recommendations, that is, determining which users are the most appropriate target audience for a given item, rather than which items should be recommended to a user.\footnote{Such a problem setting was also given in the ACM RecSys 2017 Challenge~\cite{Abel2017Challenge,Ludewig2017LightWeight}.} Their work addresses the Target Customer Distortion problem, where conventional recommender systems distort the true customer base of an item, recommending it to users whose profiles do not align with the item’s actual audience based on historical data. To tackle this, the authors apply calibrated recommendation techniques within this reversed recommendation framework.

\section{Evaluation of Calibrated Recommendations}
\label{sec:evaluation}

Having discussed technical approaches to calibration, we will now turn our attention on how these proposals are evaluated in the literature, focusing on application domains and datasets, experimental designs, and evaluation metrics.

\subsection{Application Domains and Datasets}
\begin{figure}[h!]
    \centering
    \includegraphics[width=0.7\linewidth]{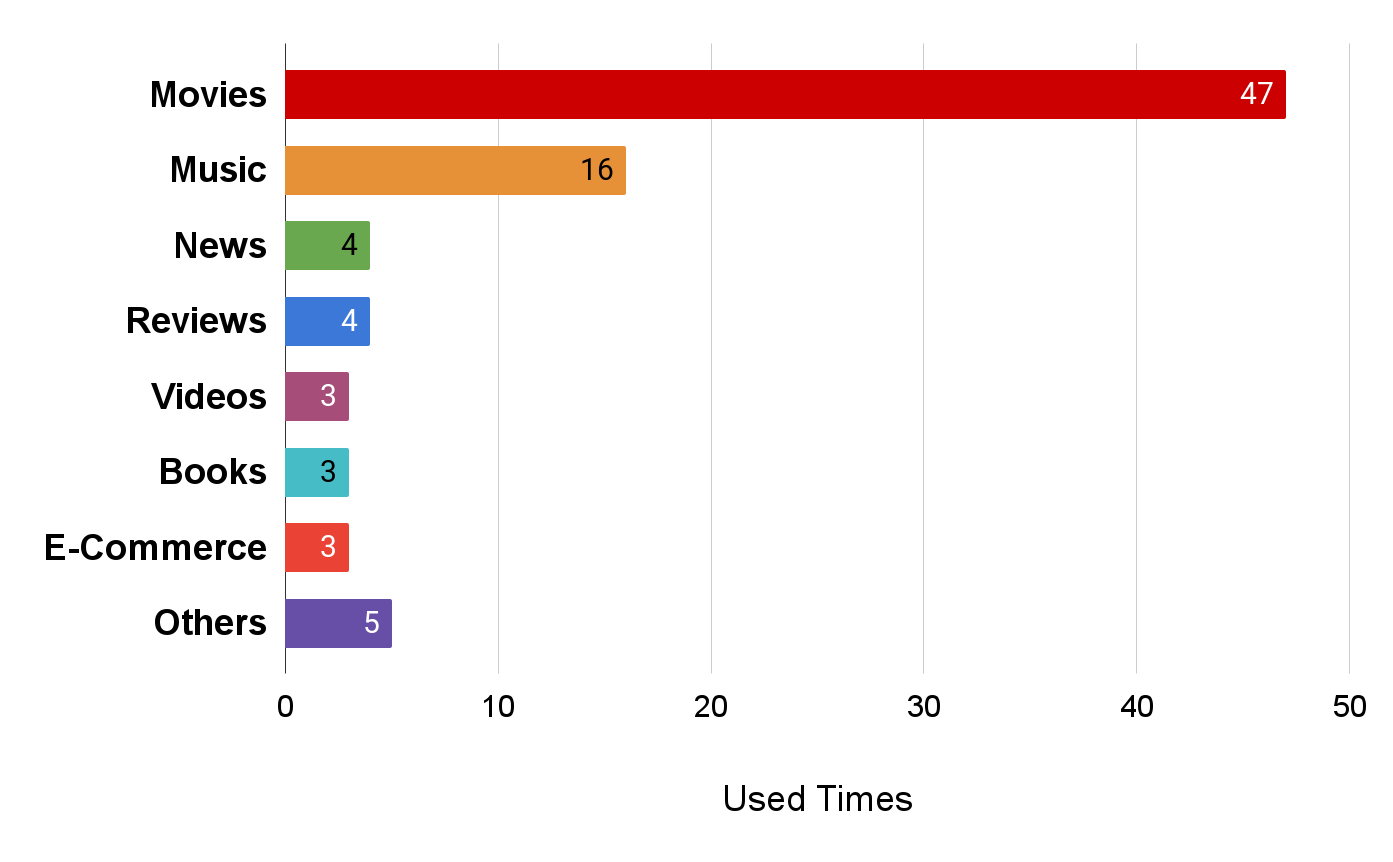}
    \caption{Number of papers per application domain.}
    \Description[Single figure]{This image is a bar chart titled ``Number of papers per application domain'', illustrating how frequently different movie-related categories appear in academic papers. The Y-axis lists the application domains, including Music (16 papers), News (4), Reviews (4), Videos (3), Books (3), E-Commerce (3), and Others (5), while the X-axis shows the frequency of use, labeled "Used Times", with values ranging from 0 to 50 in increments of 10. The tallest bar represents Music (16), making it the most referenced domain, followed by News and Reviews (4 each), while Videos, Books, and E-Commerce are the least cited (3 each). The "Others" category groups less common domains, totaling 5 papers. The caption summarizes the chart as ``Number of papers per application domain.''}
    \label{fig:domain}
\end{figure}

Figure~\ref{fig:domain} provides an overview on the studied application domains of calibrated recommendations. We observe a strong research focus on the movie domain, and almost every paper in our analysis considers a dataset from this domain in the evaluation. Among the other use cases, only music recommendation scenarios are considered to a notable extent. All other application domains are in the focus of less than five papers. The ``Others'' category in Figure~\ref{fig:domain} includes papers on food, recipes, games, podcasts, and dating.

Like in the broader recommender systems literature, studies focusing on the movie recommendation scenario seem largely over-represented, and a majority of these papers are based on one of the MovieLens datasets.\footnote{See the study by Zangerle and Bauer~\cite{Zangerle:2022:EvaluatingRS}, who also observe a dominance of MovieLens datasets in offline evaluations.} Conversely, other domains remain underexplored, highlighting the need for greater diversity in evaluation datasets.  This gap presents an opportunity for future research to expand calibration techniques into less-explored domains and compare their effectiveness with well-established areas like movies and music.

\subsection{Evaluation Methodologies}
\begin{figure*}[!ht]
    \centering
        \includegraphics[width=.7\linewidth]{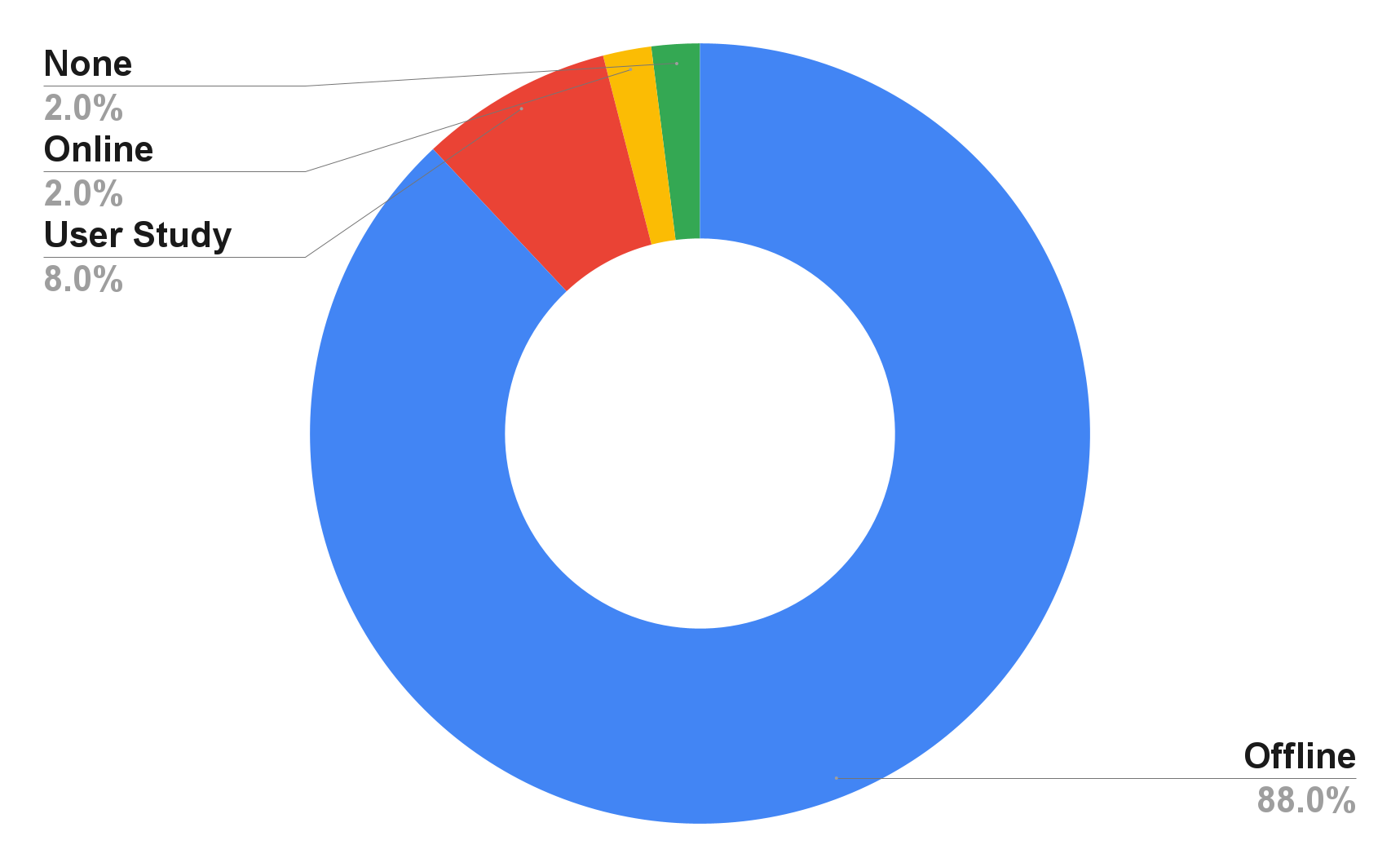}
    \caption{Distribution of evaluation methodologies.}
    \Description[Single figure]{This image presents a pie chart or percentage breakdown titled "Distribution of evaluation methodologies", showing the relative prevalence of different assessment approaches in a study or research context. The largest segment represents Offline evaluation at 88.0\%, indicating it is the dominant methodology. Smaller portions include User Study (8.0\%), and both None and Online at 2.0\% each. The visualization highlights a strong preference for offline methods over user studies and online or unspecified approaches, with precise percentages provided for each category. The caption summarizes the content as "Distribution of evaluation methodologies."}
    \label{fig:evaluation-methodologies}
\end{figure*}

Three primary evaluation methodologies are commonly used to assess the efficacy of recommendation systems: offline evaluation, user studies, and online experiments~\cite{Zangerle:2022:EvaluatingRS}. Figure~\ref{fig:evaluation-methodologies} shows the distribution of the chosen evaluation approaches in the surveyed papers.

Offline evaluation approaches strongly dominate the research landscape in the area of calibrated recommendations. This aligns well with observations in other areas of recommender systems research, as reported in other recent surveys~\cite{Klimashevskaia2024Survey,dejdjoo2023fairness}. Specifically, in our survey, we found that 88\% of the analyzed studies employ an offline evaluation protocol, while only 8\% conduct user studies, and 2\%, i.e., one paper, perform online experiments. One paper does not provide any experimental results. Generally, although offline evaluations allow for extensive benchmarking on multiple datasets and with multiple accuracy and beyond-accuracy metrics, the limited number of studies with humans strongly restricts our understanding of how calibration impacts real user preferences and behaviors. Thus, our study points to a major research gap in research on calibrated recommendations.

\subsection{Offline Experimental Designs}

Most papers rely on established offline evaluation setups, where the existing data is split into training, validation, and test data, and the splitting is either done randomly or based on time. Commonly, two or more datasets are used for the evaluation.\footnote{An exception is Steck's original paper, where only one single MovieLens dataset was used and where the test data split only contained 1\% of the interactions.}

\begin{figure*}[!ht]
    \centering
    \begin{subfigure}{0.47\textwidth}
        \centering
        \includegraphics[width=1\linewidth]{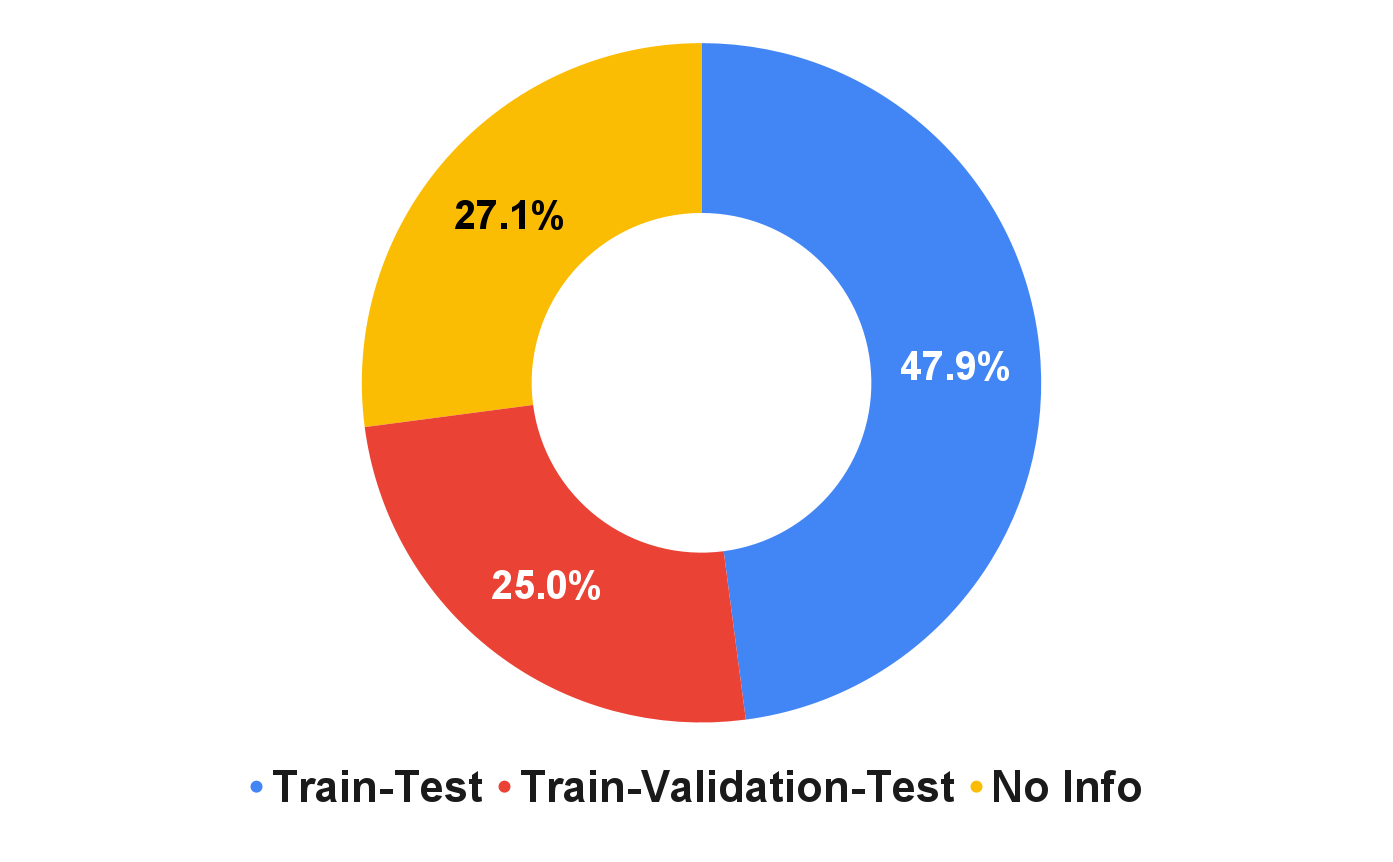}
        \caption{Data splitting partitions.}
        \label{fig:methodology}
    \end{subfigure}
    \begin{subfigure}{0.47\textwidth}
        \centering
        \includegraphics[width=1\linewidth]{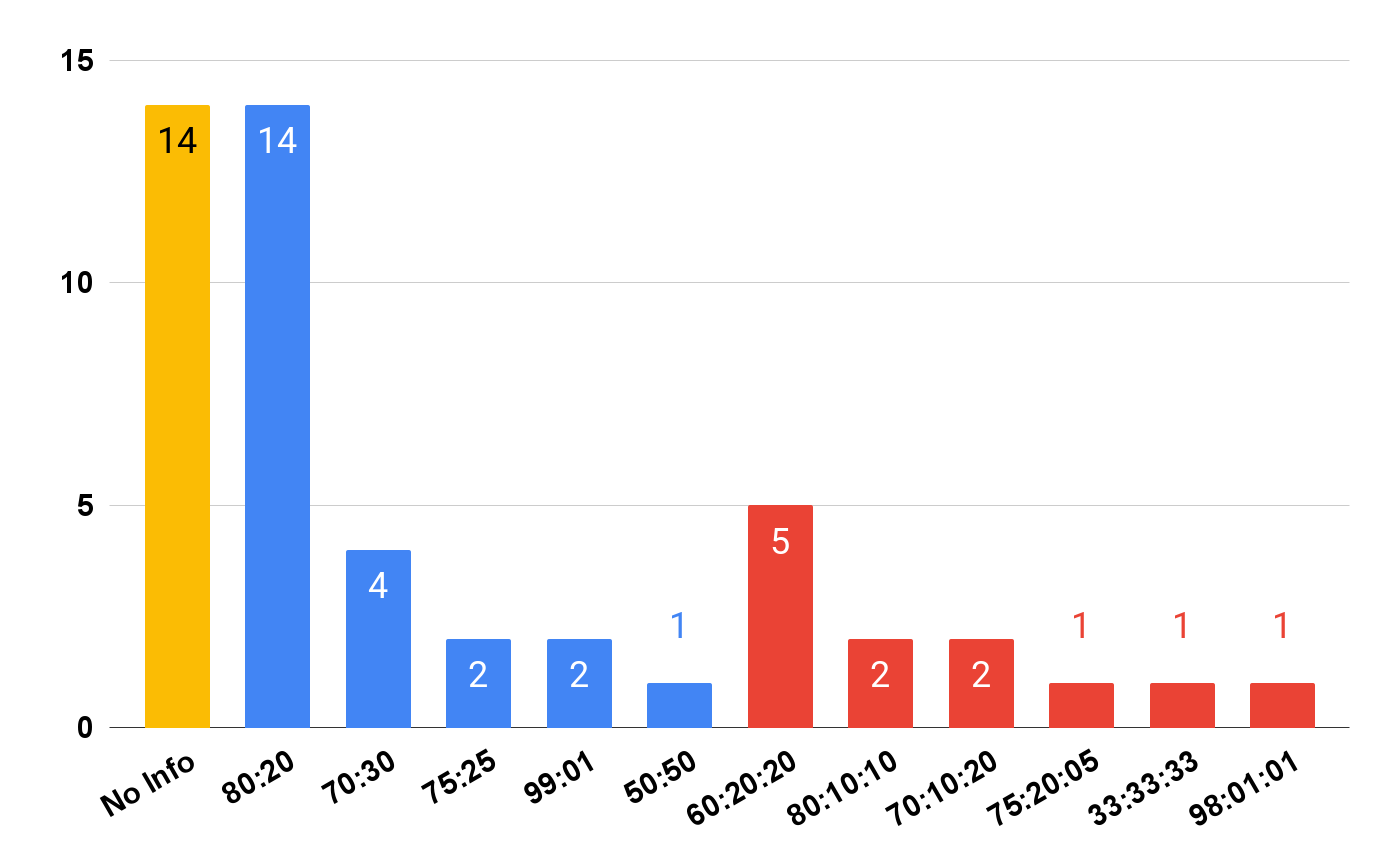}
        \caption{Data partitioning ratios.}
        \label{fig:splitnumbers}
    \end{subfigure}

    \caption{Data splitting overview, splitting partitions and ratios.}
    \Description[Two figures]{This combined description covers both figures, titled ``Data splitting partitions'' and ``Data partitioning ratios'', which analyze how datasets are used among the studies. The first figure shows that 47.9\% of studies use Train-Validation-Test splits, 27.1\% use Train-Test, and 25.0\% provide No Information. The second figure details specific split ratios and their frequencies: the most common is 80:20 (15 studies), followed by 70:30 and 75:25 (14 each), while rarer splits include 50:50 (10), 60:20:20 (5), and extreme cases like 99:01 (4) and 98:01:01 (1). Multi-way splits such as 33:33:33 (1) also appear. Together, these reveal a preference for standard splits, though methodologies vary significantly, with a quarter of studies omitting partition details entirely. Captions: "Data splitting partitions" and "Data partitioning ratios." (Visual elements like colors or chart types are omitted for accessibility.)}
    \label{fig:splitdataset}
\end{figure*}

Figure~\ref{fig:splitdataset} provides an overview of the data splitting strategies used in the considered studies. Worryingly, we find that 27.1\% of the papers do not report any information regarding their data partitioning, which limits reproducibility. Among those that do, 25\% adopt a three-way split into training, validation, and test sets. According to Figure~\ref{fig:splitnumbers}, the most common configuration for this setup is 60\% for training, 20\% for validation, and 20\% for testing (60:20:20). About half of the studies (47.9\%) use a simpler split into training and test sets, with the most frequent ratio being 80\% for training and 20\% for testing (80:20). These findings suggest that a large number of papers may rely on a weaker offline evaluation protocol using only training and test sets, rather than the more robust three-way split that includes a validation phase.

\paragraph{Baseline Algorithms}
The majority of the papers apply calibration as a post-processing step, as discussed above. Thus, arbitrary baseline rankers can be used in these cases. Considering the \emph{family} of the used baseline ranking techniques, we find that \emph{collaborative filtering} (CF) is the dominant approach and almost 90\% of the analyzed articles report experiments with a CF-based algorithm. Experiments using \emph{content-based filtering} techniques are rather rare, and can only be found in about 10\% of the studies.  Other recommendation techniques (e.g., context-aware and hybrid models) have yet to be explored within the field of calibration.

\begin{figure}[h!]
    \centering
    \includegraphics[width=0.7\linewidth]{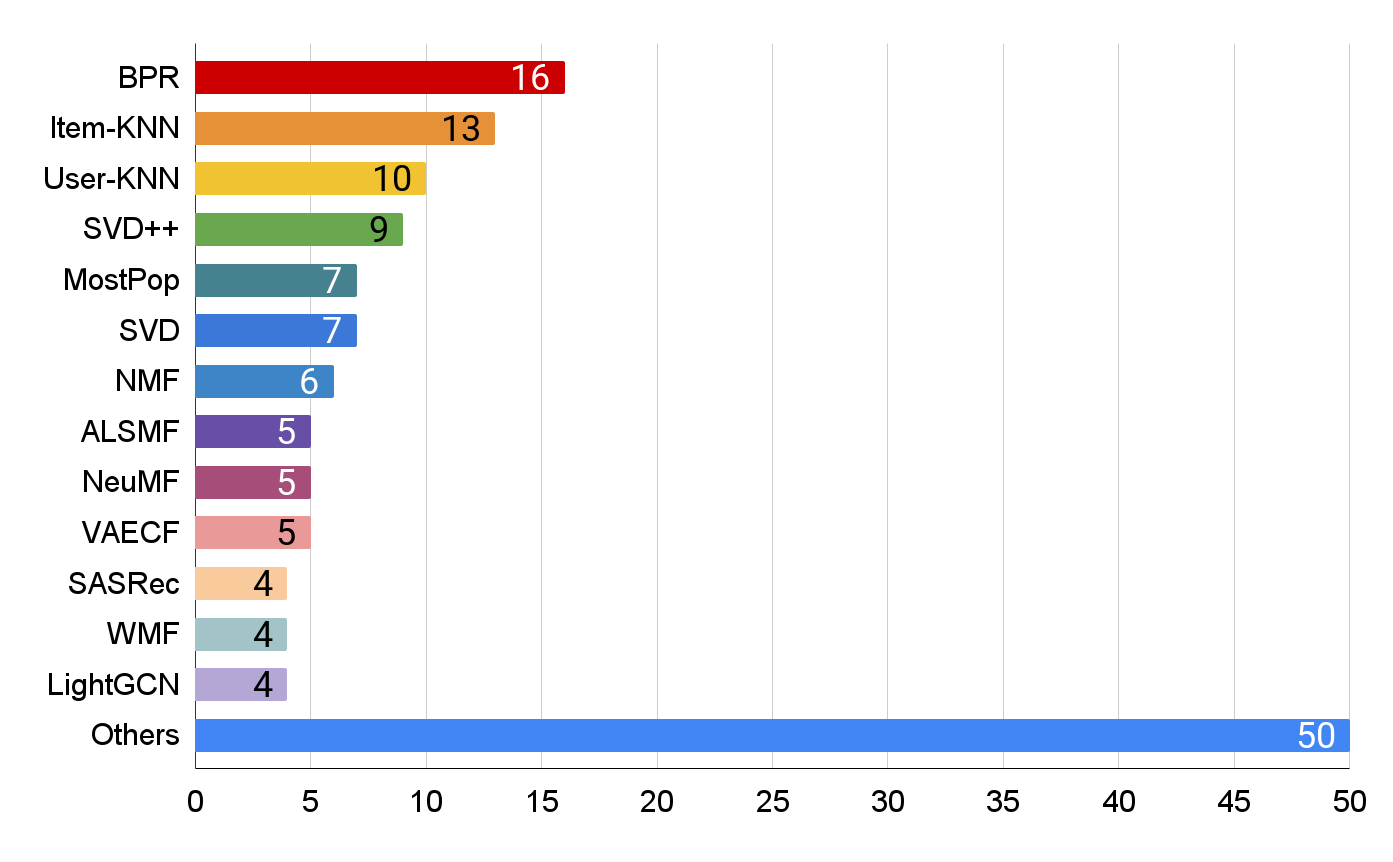}
    \caption{Frequency distribution of baseline algorithms. The ``Others'' category comprises 38 algorithms that were used fewer than three times.}
    \Description[Single figure]{This bar chart, titled ``Frequency distribution of baseline algorithms'', shows how often different recommender system algorithms were used as baselines in research studies. The Y-axis lists algorithms by name with their usage counts, while the X-axis shows the frequency range from 0 to 50. BPR is the most common (16 studies), followed by Item-KNN (13) and User-KNN (10). Mid-range frequencies include SVD++ (9), MostPop and SVD (7 each), and NMF (6). Less frequent algorithms (≤5 uses) include ALSMF, NeuMF, VAECF, SASRec, WMF, and LightGCN. The "Others" category aggregates 38 rarely used algorithms (each appearing fewer than 3 times). The chart highlights the dominance of classic collaborative filtering methods (e.g., BPR, KNN variants) over newer or niche approaches.}
    \label{fig:baseline_rankers}
\end{figure}

In terms of CF-based approaches, a rich variety of techniques were considered in the literature. The three most frequently used individual models are BPR~\cite{Steffen:2012:BPR}, ItemKNN~\cite{Sarwar:2001:ICF:371920.372071} and UserKNN~\cite{Resnick:1994:GOA:192844.192905} (Figure \ref{fig:baseline_rankers}). Across all surveyed papers, more than two dozen other algorithms were considered in the experiment, including both various matrix factorization models and deep learning models. Experiments were also made both for sequential recommendation settings as well as for traditional \emph{top-n} recommendation scenarios.

\paragraph{Calibration Targets} In most studies, item genres are used as the target category/feature for deriving user distributions~\cite{Steck:2018:Calib, Silva:2021:Exploiting, Silva:2023:Evaluating, Starychfojtu:2020:SmartRecepies, Seymen:2021:Constraint, Chen:2023:Triple, Zhao:2024:Dating}. However, alternative categorical attributes are used as well. For instance, Abdollahpouri et al.~\cite{Abdollahpouri:2021:UserCentered} introduce Calibration Popularity (CP), a variation of Steck's original method that replaces genres with item popularity levels. In this approach, items are categorized into three groups (Head, Middle, and Tail) based on their popularity. The user's preference distribution is then derived from the proportions of interactions with items from each of these groups. Following Abdollahpouri, several other studies~\cite{Klimashevskaia:2022:Mitigating, Lesota:2022:Cross_Group, Sacilotti:2023:Popularity, Klimashevskaia:2023:Evaluating, Souza:2024:TwoStage, Ungruh:2024:Mittigation, Souza:2024:Popularity, Jiayi:2023:LongTail} focus on item popularity. In~\cite{Zhao:2020:Distortion}, in contrast, the user's gender and age are used as categories to derive the distribution, because in this study, users are recommended for items. Wang et al.~\cite{Wang:2023:Quality} apply a quality-aware approach to calibration, where the quality of the items can be \emph{high} or \emph{normal}. Lin et al.~\cite{Lin:2024:Dynamics} use the video tag manually set by a specialist as a feature. Also inspired by work on intent-aware recommendation, Kaya and Bridge~\cite{Kaya:2019:Intent} instead of item features, they implement an idea called \emph{user subprofile} as a feature, where a subprofile is a set of items that capture one of the user's interests. The authors combine this subprofile idea with calibrated recommendations based on their previous studies, where they develop a method called Subprofile-Aware Diversification (SPAD).

\paragraph{Metrics}
In terms of evaluation metrics, we recall that calibration is considered a trade-off problem between accuracy and other quality factors. Thus, every paper in our survey that relies on offline evaluation uses at least one metric to assess recommendation accuracy. The three most popular list-based metrics are NDCG, Precision, and Recall. Other metrics such as MRR or MAP are used less frequently; the prediction error metric MAE is considered in one single paper~\cite{Kowald:2023:Study}.

\begin{figure}[h!]
    \centering
    \includegraphics[width=0.7\linewidth]{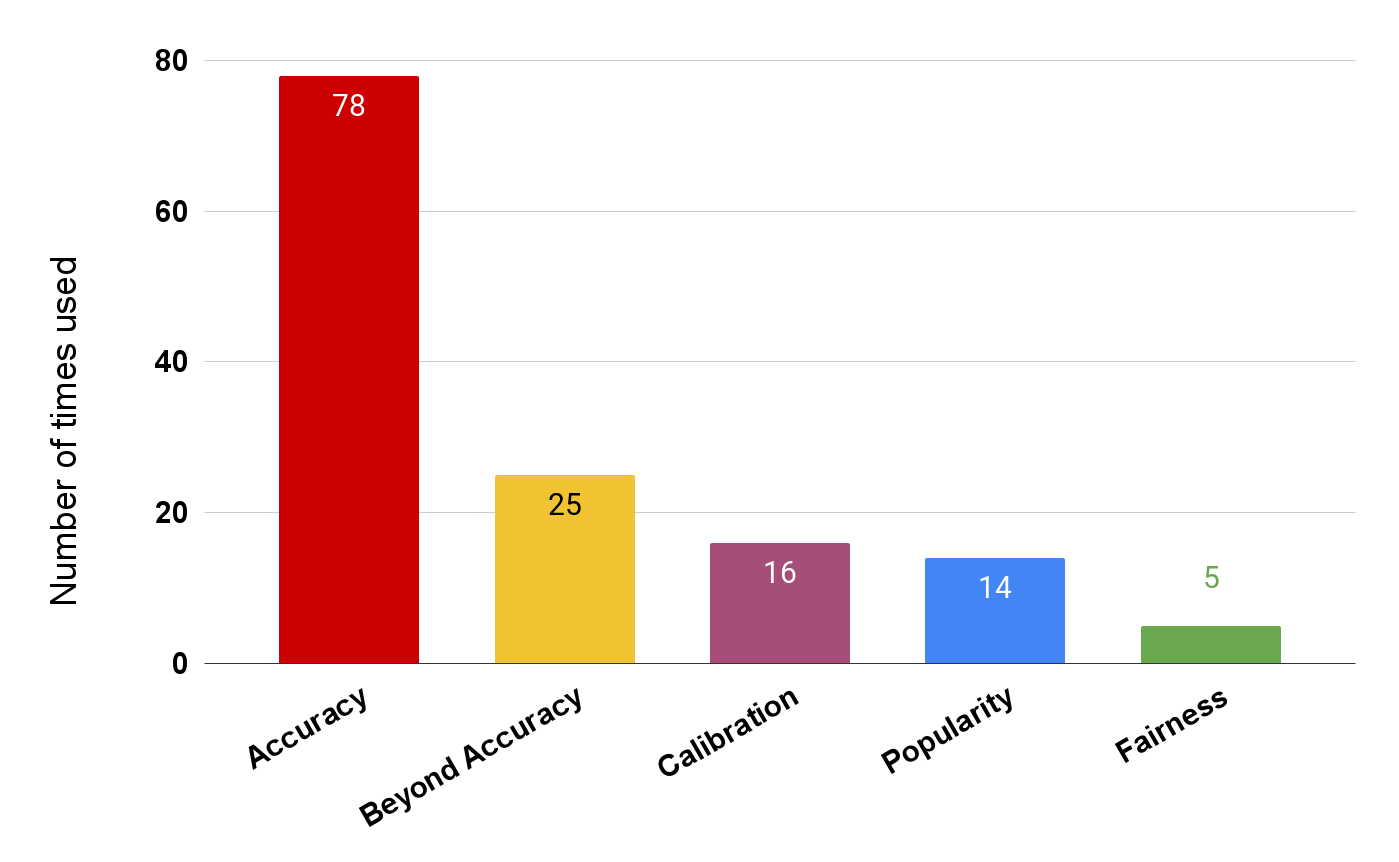}
    \caption{Focus of the calibration process in offline evaluations.}
    \Description[Single figure]{This bar chart, titled "Focus of the calibration process in offline evaluations", displays how frequently different evaluation metrics were used in studies. The Y-axis represents the "Number of times used", ranging from 0 to 80, while the X-axis lists the metric categories. Accuracy is the most dominant metric (78 uses), followed by Beyond Accuracy (25), Calibration (16), Popularity (14), and Fairness (5). The chart highlights a strong emphasis on traditional accuracy measurements.}
    \label{fig:metrics}
\end{figure}

In terms of other quality factors, a rich variety of metrics is used. We group them into the following categories: \emph{Beyond Accuracy}, \emph{Popularity}, and \emph{Fairness}. An overview of the use of different metrics and the corresponding optimization is shown in Figure~\ref{fig:metrics}.

The group of \emph{Beyond Accuracy} metrics include aspects such as recommendation novelty, as studied in~\cite{Silva:2023:Evaluating, Naghiaei:2024:BeyondAccuracy, Gomez:2024:AMBAR}, diversity~\cite{Kaya:2019:Intent, Abdollahpouri:2021:UserCentered, Naghiaei:2022:Towards, Treuillier:2024:ADF}, coverage~\cite{Silva:2023:Evaluating, Abdollahpouri:2021:UserCentered, Seymen:2021:Constraint, Abdollahpouri:2021:Target, Naghiaei:2022:Towards, Nazari:2022:Podcast, Wang:2023:Quality, Jiayi:2023:LongTail, Sacilotti:2023:Popularity, Naghiaei:2024:BeyondAccuracy, Gomez:2024:AMBAR, Ortega:2024:LLM, Souza:2024:Popularity}, or unexpectedness and serendipity~\cite{Silva:2023:Evaluating}. The tendency of algorithms to focus on popular items before and after calibration (\emph{Popularity}) is assessed in a variety of ways in the studied literature, including Average Recommendation Popularity (ARP)~\cite{Klimashevskaia:2022:Mitigating, Naghiaei:2024:BeyondAccuracy, Seymen:2021:Constraint}, Group Average Popularity (GAP)~\cite{Abdollahpouri:2020:PopularityCalibration, Souza:2024:TwoStage}, User and Item Popularity Deviation~\cite{Abdollahpouri:2021:UserCentered, Klimashevskaia:2022:Mitigating}, Popularity Lift~\cite{Kowald:2023:Study}, Bias Disparity~\cite{Gomez:20024:MOREGIN}, Average Percentage of Long Tail Items (APLT) and Average Coverage of Long Tail items (ACLT)~\cite{Klimashevskaia:2022:Mitigating}. A smaller number of works aim to assess the \emph{Fairness} of the recommendations through corresponding metrics, e.g.~\cite{Dimitris:2019:Common,Seymen:2021:Constraint}.

Generally, papers that are based on offline evaluations\footnote{Papers that assess quality perceptions via user studies not necessarily have to quantify the level of miscalibration, but rather target at directly assessing recommendation quality.} implement some metric to quantify the level of miscalibration. Various ways are used in the literature to measure miscalibration, e.g., with the help of the Kullback-Leibler divergence~\cite{Steck:2018:Calib}, the Earth-Mover's Distance~\cite{Oh:2011:Novel}, or the Jensen-Shannon divergence~\cite{Abdollahpouri:2021:UserCentered}, as discussed above. Besides these general methods to quantify the divergence of two distributions, a number of calibration-specific measures were proposed in the surveyed literature.

\citet{Silva:2021:Exploiting}, for example, propose the \emph{Mean Average Calibration Error (MACE)} and the \emph{Mean Rank Miscalibration (MRMC)} metrics. These normalized metrics were designed to enable a valid comparison of existing calibration approaches from the literature which were originally evaluated using different miscalibration measures, e.g., when one paper used the Kullback-Leibler divergence and the other relied on the Jensen-Shannon divergence. In a subsequent related work, \citet{Silva:2023:Protocol} introduce two more metrics, \emph{Coefficient of Miscalibration (CMC)} and \emph{Coefficient of Calibration Error (CCE)}. These metrics represent further developments of the MACE and MRMC metrics, which consider the Mean Average Precision (MAP) in the equation, aiming to ensure that calibration quality is evaluated relative to ranking accuracy. Lower metric (error) values indicate a desirable balance between high MAP values and low miscalibration scores, representing an optimal trade-off. Finally, \citet{Chen:2023:Triple} introduce the \emph{Neighbor Average Calibration Error (NACE)} and \emph{Neighbor Average Genre Rating Calibration Error (NAGRCE)} metrics. Drawing on inspiration from~\cite{Silva:2021:Exploiting}, these metrics were designed for use in a neighborhood-based approach to calibration. NACE measures the average mismatch between the user's target distribution and those of neighboring users, assessing how well the recommendations align within a local user cluster. NAGRCE extends this concept by capturing differences in genre-based ratings, comparing the target user's ratings across genres with those of their neighbors.

\subsection{User Studies and Field Tests}
Although offline evaluations remain the most widely used approach in recommender system research, as noted by Zangerle et al.~\cite{Zangerle:2022:EvaluatingRS}, they primarily offer a quick, reproducible, and controlled evaluation of system performance using well-defined metrics. Evaluations with humans, in contrast, can provide a reliable understanding of how real users perceive and interact with calibrated recommendations. Such evaluations with humans are commonly either implemented as controlled \emph{user studies}, where study participants typically interact with a prototype system that was developed for the purpose of the study, or as \emph{field tests}, where different versions of a system are tested in a real-world system in the form of an A/B test. In the case of A/B tests, a common approach is to first conduct offline experiments, select the most promising algorithms, and then validate their effectiveness online. Such a setup ensures that a calibration model not only performs well in controlled offline settings, but also translates to improved user experience in real-world applications.

\paragraph{User Studies}
Our survey surfaced only five papers that involve studies with humans. Four of them report outcome of user studies~\cite{Starychfojtu:2020:SmartRecipies, Lesota:2023:Perceived, Alves:2024:UserPerception, Ungruh:2024:Mittigation}.
Starychfojtu et al.~\cite{Starychfojtu:2020:SmartRecipies} address the problem of providing users with inspiration through recommendations in the context of cooking and food shopping. They develop a mobile app that generates recommendations based on past shopping lists of users, and they conduct a preliminary user study to assess the quality perceptions of three recommendation strategies: similarity-based, diversity-enhanced, and calibration-based. 32 respondents participated in the study and provided numerical ratings for recipe recommendations that were based on a set of food items in a given shopping cart. The results showed that pure similarity-based recommendations were perceived as being significantly worse than the diversity-enhanced and calibration-based ones. The calibrated recommendations were found to be slightly better than the diversity-enhanced ones, which can be attributed to the more sophisticated calibration approach based on the Fuzzy D'Hondt's algorithm that was implemented for the study.

In a more recent work, Alves et al.~\cite{Alves:2024:UserPerception} aim to assess user perceptions when the system's recommendations are calibrated for \emph{fairness}. In their study in the movie domain, the authors consider two situations that may be considered to lead to unfair situations. First, it might be considered unfair if a system mostly recommends already popular movies, reinforcing the system's popularity bias~\cite{Klimashevskaia2024Survey}. Second, one may consider it unfair if low-budget productions do not receive enough exposure by the system. For the user study, corresponding calibration techniques were implemented that ultimately lead to a more frequent recommendation of movies that are not blockbusters and were produced with a limited budget. An online study with 500 participants showed that a properly tuned calibration technique can be effective in guiding the choices of the participants without leading to a decreased quality perception of the recommendations. Notably, Alves et al.~\cite{Alves:2024:UserPerception} also observe that users only report noticing fairness improvements when explicitly alerted by the system through an explanation. This suggests that users should be explicitly signaled about this aspect within the user interface.

The problem of popularity bias in system-generated recommendations is also the focus of the study by Lesota et al.~\cite{Lesota:2023:Perceived}. The authors conducted an online study in the music domain involving 56 participants. In their within-subjects experiment, the participants were shown five personalized recommendation lists of ten items, where each of them was created with a different algorithm. Among other aspects, the participants were then tasked to provide an assessment of the popularity bias of the list. These assessments were then correlated with the level of \emph{popularity miscalibration}, as measured through the Jensen-Shannon Distance, to analyze if this computational measure can be used as a proxy for user perceptions. Differently from the findings reported earlier by \cite{Ferwerda2023Idontcare} on perceived popularity bias, Lesota et al.~found \emph{``that users generally
\emph{do} perceive significant differences in terms of popularity bias between algorithms if this bias is framed as popularity miscalibration.''}

Popularity-calibrated recommendations in the music domain were also the focus of the user study by Ungruh et al.~\cite{Ungruh:2024:Mittigation}. In their study, the authors---like in the work by Alves et al.~\cite{Alves:2024:UserPerception}---aimed to assess user perceptions when presented with calibrated recommendations. They recruited 40 participants with an active Spotify account, who then interacted with a tool that was developed for the study, in which they completed listening sessions and were then asked to choose some of the recommended items to add to their playlist. Different recommendation algorithms and calibration techniques were explored in the study, which produced recommendations that differed in their popularity bias and which measurably influenced the choices of the participants. In terms of user perceptions, the authors found that the varying levels of popularity bias did \emph{not} significantly affect the users' satisfaction with the recommendations. For some approaches, an effect on the users' \emph{familiarity} with the recommendations was observed. The authors argue that this represents an opportunity for discovery, which may ultimately lead to higher satisfaction by promoting less popular items. In terms of the perception of \emph{popularity bias}, Ungruh et al.~\cite{Ungruh:2024:Mittigation} found that users noticed differences in popularity only when using one of the two explored calibration techniques. Overall, combining the results by Lesota et al.~\cite{Lesota:2023:Perceived} and Ungruh et al.~\cite{Ungruh:2024:Mittigation} on user perceptions of popularity changes indicates that more research and studies involving more participants seem required in this area in the future.

\paragraph{Field Tests}
We could only find one single paper that reports on the outcome of field-testing a calibration approach. Klimashevskaia et al.~\cite{Klimashevskaia:2023:Evaluating} implement the Calibrated Popularity (CP) technique in a production system, evaluating its impact in a real-world setting. Their findings of an A/B test, which involved calibrated and non-calibrated recommendations, indicate that calibration can be used to effectively promote less-popular items without negatively impacting relevance, as measured through the Click-Through-Rate (CTR). This suggests that the common accuracy-calibration trade-off may not necessarily exist in real-world settings. The study however also revealed practical challenges when applying calibration in a real-world settings. One particular challenge lies in the question how to set a proper threshold that is used to distinguish popular from less popular content. Depending on the threshold setting, the ``intervention'' through the calibration technique may for example be too limited to lead to a measurable effect on user behavior. Furthermore, since A/B tests are often only run for a limited amount of time, e.g., a few weeks, it is difficult to predict long-term effects of calibrated recommendations on user behavior and satisfaction.

\section{Discussion}
\label{sec:discussion}

Most studies indicate that calibration can enhance recommendation systems by improving key performance metrics. However, this benefit typically comes at the expense of increased computational cost, primarily due to the additional re-ranking step involved in the calibration process. The effectiveness of calibration varies depending on the system architecture and evaluation methodology, resulting in a wide range of reported outcomes. Consequently, several open challenges and research directions remain, warranting further investigation in future studies.

\subsection{Issues}
\paragraph{Challenges of Calibrating the Calibration Process (Thresholds, Weights)}
A key challenge in calibration is determining the optimal settings to maximize performance. One such challenge is selecting the appropriate threshold for the number of items used in re-ranking. A higher threshold increases the computational cost but also expands the set of candidate items, potentially improving calibration. However, no standard threshold values have been widely established, leaving this issue unresolved. Another critical factor is the trade-off weight $\lambda$, which balances calibration and accuracy. Most studies evaluate a range of values for $\lambda$ (e.g., 0.0 to 1.0), as performance fluctuates depending on the system. While personalized approaches to determine $\lambda$ have been proposed, their effectiveness remains underexplored, limiting our understanding of their applicability. Overall, optimizing the selection of both the threshold and $\lambda$ remains an open problem in the field. Further research is needed to evaluate these parameters systematically and develop adaptive strategies for improving the effectiveness of calibrated recommendations.

\paragraph{Computational Complexity and Scalability Challenges}
Calibrated recommendations are NP-hard, as generating the optimal recommendation list requires evaluating all possible combinations of candidate items across all possible positions in the list. This results in a combinatorial explosion, where the complexity increases exponentially with the number of items, genres, users, and the length of the recommendation list. Jeon et al.~\cite{Jeon:2024:LeapRec} conducted a time complexity analysis across various studies, highlighting that some calibration methods have high computational costs. Similarly, da Silva et al.~\cite{Silva:2025:TimeEntropy} compiled runtime measurements from multiple experiments, showing that the time required to derive the distribution in calibrated recommendations scales with dataset size. These findings indicate that scalability is a major concern in the field. Most studies are conducted in controlled environments, with predefined numbers of users and items. However, real-world production systems operate under dynamic, large-scale conditions where computational efficiency is crucial. The trade-off between improved performance and system responsiveness may become impractical if a calibration approach requires excessive computation time. This challenge presents a research opportunity for new methods aimed at reducing computational costs in calibration while preserving its benefits in evaluation metrics. Future work should explore efficient approximation techniques, heuristics, and parallelization strategies to improve the scalability of calibrated recommendation algorithms.

\paragraph{Limitations of Calibration, possible undesired effects in practice}
Numerous studies have shown that calibration can influence additional recommendation objectives such as coverage, diversity, popularity bias, and serendipity. These effects are generally positive, often leading to improved system performance. However, in some cases, particularly regarding accuracy, calibration has been observed to reduce performance. Beyond accuracy, other potential negative side effects of calibration remain underexplored in the literature. For example, in news recommendation systems, where recency and popularity are critical, calibration may have a negative impact by suppressing popular or trending content. Similarly, in domains with strong inherent bias, such as job-seeking platforms where candidates are recommended to companies, calibration may either help mitigate the bias or inadvertently maintain it.\footnote{Gender biases, for example, are not uncommon in job-seeking platforms~\cite{Ludewig:2017:job, li:2023:fairness, Kumar:2023:jobfairness}.}
Understanding the context and underlying data distribution is essential to assess whether calibration leads to fairer recommendations or introduces unintended consequences. Moreover, calibration introduces financial and operational implications due to increased computational complexity. In real-world deployments, the benefits of calibration may be offset by the associated production costs, raising concerns about its scalability and feasibility in practice.

\paragraph{Calibration Can Lead to a Reduction of Performance for Individual Users}
Evaluation in recommender systems typically relies on reporting the average performance across all users, resulting in a single aggregate metric. The calibration literature follows this same practice. However, such aggregated evaluations can obscure important user-level variations. For some users, the calibrated recommendation list may underperform, failing to reduce miscalibration, decreasing accuracy, or lacking novelty and diversity. As a result, the individual-level impact of calibration remains largely unexplored, highlighting a critical gap in current research.

\paragraph{User Perception}
An issue reported by Alves et al.~\cite{Alves:2024:UserPerception} and Lesota et al.~\cite{Lesota:2023:Perceived} is that users only perceive a recommendation list as calibrated or fair when it is accompanied by a label or explanatory text indicating so. This suggests that the effects of calibration may be too subtle to be noticed by users on their own, or that the calibration does not meaningfully impact the user experience. Notably, this perception gap occurs despite observed improvements in objective performance metrics, indicating that the effectiveness of calibration from a system perspective does not necessarily translate into perceived value for the user. This raises questions about how calibration should be designed or communicated to be both effective and perceptible.

\subsection{Future Work and Outlook}

\paragraph{When to calibrate}
Typically, calibration is applied uniformly across all users in the system, with the same calibration degree. However, a few studies, such as \cite{Silva:2021:Exploiting, Silva:2023:Protocol, Silva:2025:Benchmark, Wang:2021:Deconfounded, Lesota:2022:Cross_Group}, have proposed personalized approaches to adjust the calibration level for each user. Additionally, Sacilotti et al.~\cite{Sacilotti:2023:Popularity} introduce a method to personalize calibration based on the user's preferred attribute distribution (genre or popularity). Despite these efforts, an important open question remains: some users may prefer to receive popular or trending items rather than a fully calibrated list. Future research should explore when and for whom calibration is appropriate, identifying users who benefit from calibration and those who may prefer alternative recommendation strategies depending on their preferences and contextual factors.

\paragraph{Dynamic Calibration (Recent Events, Intent-Awareness)}
Calibration is typically implemented as a static method, deriving the user distribution from historical preferences, as discussed throughout this survey. However, current techniques do not adapt the distribution dynamically to account for recent changes in user behavior or emerging interests. There is limited research addressing this gap. Only one study to date, by Kaya and Bridge~\cite{Kaya:2019:Intent}, explicitly explores the connection between calibration and intent-awareness, suggesting that combining these two approaches could lead to a deeper understanding of user intent and better interpretation of evolving user trends. Additionally, da Silva et al.\cite{Silva:2025:TimeEntropy} and Lin et al.\cite{Lin:2024:Dynamics} propose calibration strategies that incorporate temporal aspects of user preferences. While these approaches introduce time-aware elements, they still lack full dynamism and real-time adaptability to user preference shifts. Future work should focus on developing fully dynamic calibration methods capable of adjusting to user intent and evolving tastes in real time.

\paragraph{Underexplored Recommendation Techniques: Hybrid, Context-Aware, and Others}
Collaborative filtering remains the most widely studied technique in the calibration literature, largely due to its ease of reproduction and the availability of numerous public datasets. This trend is evident after compiling all the papers, which illustrates the predominance of collaborative approaches in calibrated recommendation studies. However, this focus restricts the applicability and generalizability of calibration techniques. Future research should expand the scope by incorporating content-based methods and exploring hybrid or context-aware recommendation techniques. Investigating the impact of calibration within these less-explored paradigms can reveal new insights and broaden the effectiveness of calibrated recommendations across diverse system architectures.

\paragraph{New domain exploration must happen}
The movie and music domains are the two most extensively explored areas in calibrated recommendation research. These domains benefit from the availability of rich datasets, such as MovieLens and Last.fm, which have become standard benchmarks in the field. In contrast, other domains remain either underexplored or entirely unexplored, with insufficient empirical results to support domain-specific inferences. This imbalance highlights an important research gap, signaling the need to expand calibration studies into diverse domains to evaluate their generalizability and effectiveness in different recommendation contexts.

\paragraph{More Field Studies}
We analyzed \finalNbOfPapers papers on calibrated recommendations. Most of these studies rely on offline evaluations, with only a few incorporating user or field studies. Notably, Klimashevskaia et al.~\cite{Klimashevskaia:2023:Evaluating} stands out as the only work that applied calibration within a real-world production recommendation system. The limited application of calibration in practical environments restricts our understanding of its real-world impact on users. Therefore, future research must advance into field studies, enabling the community to gather more practical insights and assess how calibration is perceived and experienced by real users.

\paragraph{Cold Start Problem}
As discussed earlier in this survey, calibration relies on past user interactions to estimate the user’s preference distribution. However, a significant challenge arises in cold-start scenarios, where a user has little to no historical interaction data. In such cases, it remains unclear how calibration techniques should behave and how well they can generate a balanced recommendation list while simultaneously allowing the user to explore and define their preferences.

\paragraph{Benchmarking All Proposals}
Approximately 58\% of the \finalNbOfPapers papers analyzed present a technical contribution to the field (Figure~\ref{fig:approach}). Each proposal represents an advancement, introducing new strategies to generate calibrated recommendation lists. However, no study to date has comprehensively compared all—or even most—of these proposals within a unified evaluation framework. Future research should focus on benchmarking these approaches under consistent conditions, using the same dataset, preprocessing pipeline, and baseline recommender algorithm. This would enable a fair comparison across various performance metrics, such as precision, coverage, miscalibration, diversity, and popularity. Additionally, the reproducibility of these methods must be considered, especially given the growing concern within the recommender systems community regarding the replicability of published results.

\section{Summary}
\label{sec:summary}
In this survey, we analyzed 356 papers and selected \finalNbOfPapers that specifically address the field of calibrated recommendations, plus two from previous similar ideas. Our analysis reveals a consistent growth in the number of published works over recent years, with the majority appearing in conference proceedings. Most of these contributions present technical proposals, complemented by studies focused on impact analysis and system comparisons. Calibrated recommendation systems are generally structured into several key components, and their specific implementations vary depending on system objectives, ranging from genre or popularity-based distribution derivations to methods that integrate calibration either directly into the recommender algorithm or as a post-processing step. The field is predominantly explored in movie and music domains, which offer valuable insights into real-world applicability. However, the limited number of studies involving online systems highlights a significant gap in user-centered evaluation and practical validation. This shortcoming presents opportunities for future research, many of which have been outlined in this survey. Overall, calibrated recommendation remains a relatively recent but increasingly influential area of research, with strong potential to enhance both user satisfaction and system performance.

\section*{Acknowledgments}
This research was funded in whole or in part by the Austrian Science Fund (FWF) 10.55776/COE12. For open access purposes, the author has applied a CC BY public copyright license to any author accepted manuscript version arising from this submission.

\bibliographystyle{ACM-Reference-Format}
\bibliography{references}

\end{document}